\documentclass{article}
\usepackage{a4wide}
\usepackage{epsfig}
\newcommand{\degrees}{\ensuremath{^{\circ}}}
\def\lsi{\raise0.3ex\hbox{$<$\kern-0.75em\raise-1.1ex\hbox{$\sim$}}}
\def\gsi{\raise0.3ex\hbox{$>$\kern-0.75em\raise-1.1ex\hbox{$\sim$}}}
\newcommand{\lsim}{\mathop{\lsi}}
\newcommand{\gsim}{\mathop{\gsi}}
\newcommand\abibitem[1]{\bibitem{#1}}
\newcommand\acite[1]{\cite{#1}}
\newcommand\text[1]{{\textrm{#1}}}
\renewcommand\vec[1]{{\bf #1}}

\title{Defect formation in the early universe}
\author{Arttu Rajantie
\thanks{E-mail: a.k.rajantie@damtp.cam.ac.uk}\\
DAMTP, CMS, 
University of Cambridge\\
Wilberforce Road, Cambridge CB3 0WA,
UK}
\date{13 November, 2003}

\begin{document}

\maketitle

\begin{abstract}
Topological defects are common in many everyday systems. In general,
they appear if a symmetry is broken at a rapid phase transition. In
this article, I explain why it is believed that 
they should have also been formed in the early universe and how that
would have happened. 
If topological defects are found, this will provide a
way to study observationally the first fractions of a second after the
Big Bang, but their apparent absence can also tell us many things
about the early universe.

\noindent DAMTP-2003-71
\end{abstract}

\section{Introduction}
\label{sect:intro}
Imagine sitting at a round dinner table, with a plate in front of
each guest and a spoon at half way between each two plates. This
setting is symmetric between left and right, and you could
therefore choose to use either the spoon on your left or the one
on your right. However, as soon as your neighbours make their
choice, this symmetry disappears, and you have to conform to that
choice. This is a classic example of {\it spontaneous symmetry
breaking} (SSB) (see Ref.~\acite{Peierls:fg}).

However, it may happen that the people on your right
choose to use the spoons on their left and vice versa, leaving you with
no spoon at all. This is an example of a {\it topological defect}.
They occur generally in systems with SSB,
because that requires a choice between several identical
possibilities, and the choice is often different in different parts
of the system.

Spontaneous symmetry breaking is a very common phenomenon in many
physical systems. In the Standard Model of particle physics, it is
in a sense responsible for the masses of the elementary particles,
although the associated symmetry is of a more abstract nature, as
it is a local {\it gauge} symmetry rather than a {\it global}
symmetry. The Standard Model actually happens to be a special case
with no topological defects, but
this, nevertheless, suggests that its extensions would generally
predict the existence of topological defects. Depending on the
details, these defects could be domain walls, cosmic strings or
magnetic monopoles.

Both in the early universe and in our dinner table example,
the symmetry was initially unbroken. It got broken when the first
guest picked up a spoon, or when the temperature of the universe
decreased below a certain critical temperature. At the dinner table,
it is easy to see that if the table is large enough, topological defects
are bound to form~\acite{Kibble:1976sj}, 
but the process of defect formation was more
complicated in the early universe. The purpose of this article is
to explain it.

We can also ask if these defects could have survived until today
and if and how they could be observed in that case. So far, we
have no observational evidence for their existence. This alone can
tell us many things about the early universe. On the other hand,
defects have not been ruled out, and if they are found, that would
have a huge impact on our understanding of the universe.

In order to draw any conclusions from the absence or possibly the
existence of defects,
it is crucial that we understand the process
of defect formation properly. Fortunately, we do not have to rely
solely on theoretical calculations, because topological defects
are also formed at phase transitions in certain condensed matter
systems such as superfluids and superconductors. This phenomenon
is theoretically very similar to its cosmological counterpart, and
we can use this analogy to do 'cosmological
experiments'~\acite{Zurek:qw}. 
On a
more fundamental level, these same experiments can be used to test
our understanding of non-equilibrium dynamics of quantum field
theories, which will be very important for particle physics.

In this article, I review the theory of defect formation at phase
transitions, emphasizing those aspects that are relevant for the early
universe. I will also discuss the importance of defects to
cosmology, but readers who want to know more about that are advised to
read Refs.~\acite{VilenkinShellard,Vachaspati:1998vc,Gangui:2003uu}. 
More discussion
on defect formation at phase transitions can be found in 
Refs.~\acite{Gill:1998,Gleiser:1998kk,Rajantie:2001ps}.

Throughout the paper, I will use the natural units
$c=\hbar=k_B=1$. This means that everything is expressed in units of
GeV (giga electron volt). Table \ref{table:units} shows how they can
be converted into SI units.

\begin{table}[t]
\center
\begin{tabular}{l|c|c}
 & Natural units & SI units \\
\hline
Energy & 1 GeV & $1.60\times 10^{-10}$ J\\
Temperature & 1 GeV & $1.16\times 10^{13}$ K\\
Mass & 1 GeV & $1.78\times 10^{-27}$ kg\\
Distance & 1 GeV$^{-1}$ & $1.97\times 10^{-16}$ m\\
Time & 1 GeV$^{-1}$ & $6.65\times 10^{-25}$ s
\end{tabular}
\caption{Conversion between natural and SI units.
\label{table:units}
}
\end{table}

\section{Scalar fields and global symmetries}
\label{sect:global}

Apart from gravity, all physics at microscopic scales is described
by quantum field theories (see Ref.~\cite{gaugebook}). 
For the moment, we will restrict our discussion to scalar fields,
which is the simplest type of quantum fields.
A scalar field is Lorentz invariant, which means that its value
does not depend on the reference frame. It simply 
has some value at each point in spacetime, and this value is
independent of the orientation and velocity of the observer. 
These values can be real numbers, but we can also think of complex
scalar fields, or vector or matrix valued scalar fields.

Because the field is Lorentz invariant, there is a large amount of
freedom in constructing scalar field theories. 
Most of this freedom is in the choice of the potential $V(\phi)$,
which gives the energy density of a constant field with value $\phi$.
For a real scalar field $\phi$,
the potential leads to the classical equation of motion
\begin{equation}
\ddot\phi-\vec\nabla^2\phi=-V'(\phi),
\label{equ:realeom}
\end{equation}
where the double dot indicates a second time derivative, and the prime 
indicates a derivative with respect to $\phi$. Apart
from the gradient term $\vec\nabla^2\phi$, Eq.~(\ref{equ:realeom})
is identical to the Newton equation for a particle in
the same potential $V$. Therefore, we can understand the
behaviour of the scalar field by thinking of a ball moving in a
one-dimensional landscape. 

In the vacuum state, the field is constant in space and time, and its
value corresponds to the
global minimum of $V(\phi)$. The value of $\phi$ in the
vacuum is known as the {\it vacuum expectation value} (vev).

Let us now imagine a potential that has reflection symmetry around
$\phi=0$. Typically, this would be a polynomial in $\phi^2$, for
instance
\begin{equation}
V(\phi)=\frac{1}{4}\lambda\left(\phi^2-v^2\right)^2.
\label{equ:scalarpot}
\end{equation}
The value of the potential does not change if we flip the sign of
$\phi$.
If $v^2$ is negative, the vacuum state is at $\phi=0$. 
However, if $v^2>0$, this
point becomes a local maximum, and two minima appear at $\phi=\pm v$. 
The field is now in the same situation as a
guest at the dinner discussed at the beginning of this article. It
will have to choose one of the two minima, and this breaks the
reflection symmetry spontaneously.

\begin{figure}[t]
\center
\epsfig{file=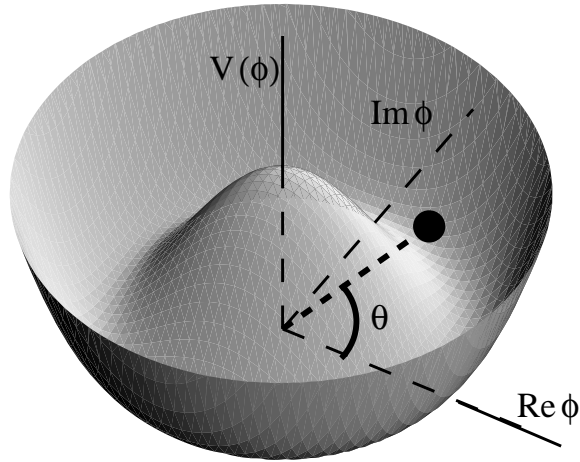,width=11cm}
\caption{
\label{fig:u1symm}
A U(1) symmetric potential for a complex scalar field. The ball
depicts the value of the field in a vacuum state at the bottom of the
potential. There is a circle of possible vacuum states, characterized
by the phase angle $\theta=\arg\phi$.
}
\end{figure}

If we have, instead of a real field,  a complex one, we can think
of a ball on a two-dimensional surface, which corresponds to the
complex plane. We will be mostly interested in the case in which
the potential depends only on the absolute value of $\phi$, not on
its phase angle. Then, this two-dimensional picture is symmetric
around the origin. A potential like this is shown in
Fig.~\ref{fig:u1symm}.
A symmetry with respect to rotations around one
axis, such as this, is known as U(1). 

We can easily think of
scalar fields with more components, although it is then more
difficult to visualize the motion. In that case, the mechanical
analogue is a ball moving in a higher-dimensional landscape, which
we will call the {\it internal space}.

Whenever a symmetry is broken, the potential has several minima, which are
related to each other by the symmetry, and each of them corresponds to a 
possible vacuum state. If the symmetry is continuous, these minima form a 
continuous valley with a perfectly flat bottom.
If we give
the ball a push in this direction, it will roll around the set of possible 
minima, no matter how weak the push was. This has the consequence 
that the quantum
theory has states with arbitrarily low energies, or in other words,
massless particles. These particles are known as {\it Goldstone bosons}.

Our previous dinner table example is not very helpful for
visualizing the breaking of a U(1) symmetry, because it only dealt
with a choice from two possibilities. In the case of U(1), we have
a continuous set of possible vacua. Perhaps the closest analogue
in everyday life is the time of the day. If we
did not interact with other people, it would not matter to which
time we set our watches. At any time of the day, we could choose any
possible value from the whole range of possibilities from 00:00 to
23:59. Because 24:00 is the same as 00:00, we can think of a circle of
possibilities. This can be interpreted as a U(1) symmetry.

However, our interaction with other people
breaks this symmetry.
The
whole society has to set their watches to the same time in order to
work properly, and this time
plays the role of the vacuum expectation value.
In the
UK it was chosen that the clocks are set to 12 o'clock when the
sun is at its highest above Greenwich, but in principle this choice
was arbitrary.

\section{Gauge fields and local gauge symmetries}
\label{sect:gauge}

The symmetries we have discussed so far have been global
symmetries: The system is only symmetric under rigid rotations,
where the field is rotated by the same amount everywhere. In our
time analogy, this corresponds to changing the time by the same
amount in the whole country. This is, of course, done every spring
and autumn, when the UK changes from Greenwich Mean Time (GMT) to
British Summer Time (BST) and back, and because of this symmetry,
it does not cause any problems. On the other hand, if you forget to move
your clock, you will most probably be missing appointments and
encountering other problems.

The Standard Model of particle physics actually
possesses an higher symmetry that allows rotations of the fields at
different points by different amount without changing their physical
content. A rotation like this is known as a local gauge
transformation, and the corresponding symmetry as
local gauge symmetry or gauge invariance.
It is made possible by the {\it gauge fields}, which also carry the
electromagnetic, weak and strong interactions. In this way, this
symmetry dictates many properties of these interactions.

We can use the time analogy to illustrate gauge
fields. As we noted, you cannot use a separate time convention
from the rest of the country without having problems, but
worldwide, different countries do use different time conventions.
This is made possible by an extra structure, 
namely the time zones. We know that
when we travel from the UK to, say, New York, we have to move our
watches five hours backwards, and this is analogous to what the
gauge field tells us in the case of a gauge field theory.
If we want to compare the scalar field to its value at some other point, we
have to first rotate it by the amount indicated by the gauge field.

Every March, when the UK moves clocks one hour forward and changes
to BST, we carry out something similar to a local gauge
transformation. Many countries do not move their clocks at the
same time, but that does not cause problems as long as travellers
are aware what the current time difference is on any given day.
Analogously, if we carry out a local gauge transformation in a
gauge field theory, the gauge field is changed in order to
compensate for the rotation of the scalar field.

The simplest example of a gauge field theory is the {\it Abelian Higgs
model}. It consists of a complex scalar field $\phi$ with a U(1)
symmetry and a gauge field $\vec{A}$, which is a real vector-valued
field.\footnote{Actually, the gauge field is a four-vector $A_\mu$,
but we ignore the time component here. This can be done consistently
without changing the physical content of the theory,
and is known as the temporal gauge.}
Because we cannot directly compare field values at two different
points, we have to replace gradients by
{\it covariant derivatives}
\begin{equation}
\vec{\nabla}\phi~~\rightarrow~~
\vec{D}\phi=\vec{\nabla}\phi+ie\vec{A}\phi,
\label{equ:covder}
\end{equation}
where $e$ is known as the {\it gauge coupling}.
(In non-Abelian theories, which we will discuss later, the gauge
coupling is usually denoted by $g$.)
Physically, this 
replacement gives the scalar particles an electric charge $e$, with
the gauge field playing the role of the vector potential. 
The electric field is given by $\vec{E}=-\partial\vec{A}/\partial t$,
and the magnetic field by 
$\vec{B}=\vec{\nabla}\times\vec{A}$. 

In the broken
phase, where $\phi$ has a non-zero value, the system becomes essentially
a superconductor. Any field configuration with a constant and non-zero
$\phi$ and $\vec{B}$ would have an infinite energy density.
This is known as the {\it Meissner effect} and is a characteristic
property of superconductors (see~Ref.~\acite{superconductors}). 
Real superconductors can be described
with a similar model, in which the role of
$\phi$ is played by Cooper pairs that consist of two electrons. 
In particle physics and cosmology, we are would think of the Abelian Higgs 
model as a relativistic theory and $\phi$ as a fundamental field. 
Then, the Meissner effect corresponds
to a non-zero photon mass and is also known as the {\it Higgs
mechanism}. 
In contrast to the global case, there are
no massless Goldstone bosons, because the rotation around the vacuum
manifold can always be compensated by the gauge field. It is therefore
often said that the gauge field ``eats'' the Goldstone boson.

In terms of our time zone analogy, the presence of a non-zero magnetic
field would correspond to a hypothetical situation, in which the
amount by which travellers need to move their clocks when they cross a
border is not determined globally by time zones, but is specified for
each border separately. Imagine that when you travel from the UK to
France, you are told to move you clock one hour forward. When you
travel further to Spain, you are again told to move your clock forward
by one hour. Finally, when you travel from Spain to the UK, you are
told to move your clock one hour backward. As a net effect, when you
return home, your watch would show one hour more than the clocks you
have in your home. In reality, this does not happen because each country
follows a specific time zone. To the extent that the choice of the time zone
corresponds to SSB, we can understand this as an example of the
Meissner effect.

\section{Particle physics}
\label{sect:particle}

In particle physics, we encounter more complicated gauge symmetries
than U(1). Weak and strong interactions are described by 
groups knowns as SU(2) and SU(3), correspondingly. 
These groups are called {\it non-Abelian}, because in these cases the gauge
field is matrix valued. The product of two matrices depends on the 
order, i.e., $AB\neq BA$, and in algebra, an operation 
with this propert is called non-Abelian or non-commutative.  
Nevertheless, the basic principles are the same as with the Abelian U(1) group.

The Standard Model of particle physics is based on SU(3), SU(2) and U(1) 
groups, and gives an extremely accurate description of strong,
weak and electromagnetic interactions in terms of gauge fields, 
quarks, leptons and a scalar field known
as the {\it Higgs field}.

The Higgs field is a complex two-component scalar field, which is sensitive 
to both U(1) and SU(2) groups.
In the vacuum, the Higgs field has a non-zero value ($v\approx 246$~GeV),
which breaks these groups spontaneously, but a residual U(1) subgroup 
survives. This U(1) group gives rise 
to electrodynamics in the same way as in the Abelian Higgs model. 
As a consequence,
the Higgs mechanism makes the W and Z bosons massive, but leaves the
photon massless.

The Standard Model has been highly successful experimentally, and in
fact, there are still no experimental results that disagree with
it. Nevertheless, it is in a sense unnecessarily complicated, because
it has three separate gauge groups, each with its own coupling
constant, and the way the quarks and leptons transform
under gauge transformations does not seem very natural. In fact, if
one calculates how the theory behaves at high energies, one finds that
at around $10^{16}$~GeV~\acite{Georgi:sy}, 
the strong, weak and electromagnetic interactions
become equally strong. Furthermore, if one thinks of these gauge
groups as subgroups of a larger gauge group, it
turns out that it is also possible to explain the gauge
transformation properties of quarks and leptons.
There are several possible ways of doing this, and they are known as
{\it grand unified theories} (GUT). The simplest example is based on
a group known as SU(5).

In GUTs, the grand unified gauge group is split into the SU(3), SU(2)
and U(1) subgroups by SSB. In the SU(5) theory, that is achieved by
postulating a new, 24-component scalar field $\Phi$, which has a
non-zero vev (of the order $10^{16}$~GeV). Unfortunately these energies
are far beyond the reach of any experiments. There are other GUTs in
which the U(1) group appears only at lower energies, such as the
SO(10) theory with the symmetry breaking pattern
\begin{eqnarray}
\text{SO(10)}&\rightarrow& 
\text{SU(4)}\times\text{SU(2)}\times\text{SU(2)}
\nonumber\\&
\rightarrow&
\text{SU(3)}\times\text{SU(2)}\times\text{U(1)},
\end{eqnarray}
where the energy scale of the latter symmetry breaking could be of the
order $10^{12}$~GeV~\acite{Kephart:2001ix}.

\section{Topological defects}
\label{sect:defects}

So far, we have been mainly discussing the vacuum states, but it is
often also possible to find other time-independent states.
Topological defects are one class of these, and
they arise in models with spontaneously broken symmetries, if the
choice of vacuum is different in different directions and these
choices do not match completely.

We already encountered one very
simple example of a topological defect in the introduction.
There, the defect arose because the symmetry breaking involved a
choice from a discrete set of possibilities. Defects like these are
called {\it domain walls}, because in three dimensions, such a defect
would form a surface separating regions of different vacua. For a real
scalar field with the potential in Eq.~(\ref{equ:scalarpot}) the
corresponding solution is the ``kink'',
\begin{equation}
\phi(x,y,z)=v\tanh\sqrt{\lambda}vz.
\label{equ:kink}
\end{equation}
This solution has a domain wall at $z=0$, where the field vanishes and
the symmetry is restored. On different sides of
the wall, the fields approach different vacua asymptotically. The kink
is stable, because it is not possible to deform the field into the
vacuum configuration without changing its value everywhere in 
half of the space simultaneously. However, if a kink
encounters an anti-kink, they annihilate.

In the case of a U(1) symmetry, the set of possible vacua, i.e. the
{\it vacuum manifold} ${\cal M}$, is a circle (see Fig.~1). 
Let us now consider 
a closed curve, which we will denote by $C$, in space. At each point on the 
curve $C$, the field $\phi$ has some value, which we can think of the position
of a ball in the potential in Fig.~1. Let us now move along the curve $C$ and
ask how the ball moves in the potential. 

In vacuum, $\phi$ has the same value 
everywhere, and the ball would therefore simply stay still in the same place,
but this is not always true.
In any case, when we have 
moved around the whole curve $C$ and come back to our starting point, the ball
must also have returned to its original place, but along the way, it may have
travelled around the circle of minima. The number of times it moved around 
the circle is known as the {\it winding number}.

Let us now assume that this winding number is not zero.
If we now deform the curve $C$ continuously, the path of the ball in the 
potential must also change continuously. If the ball stays on the vacuum 
manifold, the winding number cannot change, because that would require a 
discontinuous change. On the other hand, if we shrink the curve $C$ to a 
point, the path can consist of one point only and therefore must have winding
number zero. The only way to avoid a contradiction is that the path of the
ball has left the vacuum manifold at one point, but this means that somewhere
inside the curve $C$, there must have been a point at which $\phi$ was not 
in vacuum.
This point is known as a
{\it vortex}. 

\begin{figure}[t]
\center
\epsfig{file=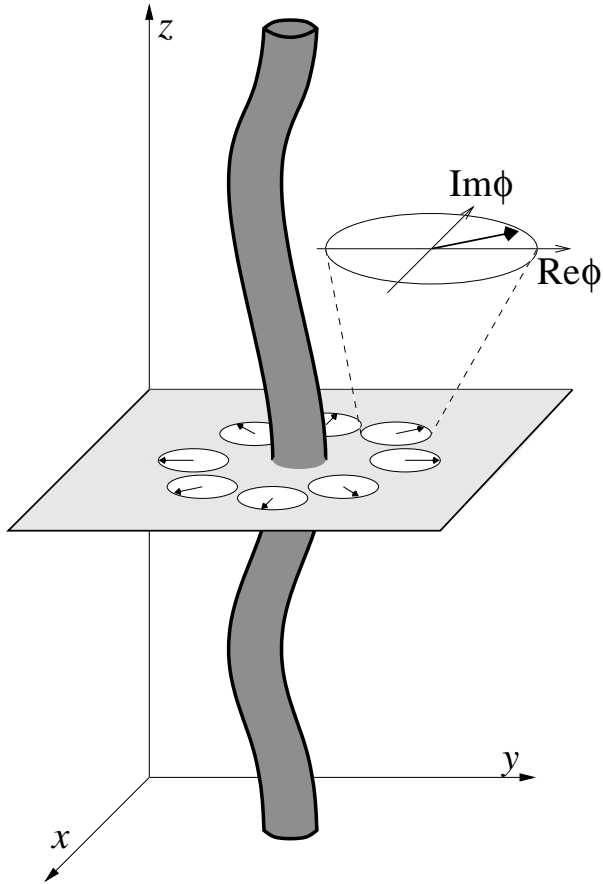,width=8cm}
\caption{
\label{fig:string}
A cosmic string, or a vortex line, in three dimensions. The direction
of the complex scalar field $\phi$, indicated by the small arrows on
the cross-sectional plane, rotates by a full 360\degrees around the
vortex. The field vanishes in the vortex core.
}
\end{figure}

In circular cylindrical coordinates, a typical vortex has the
rotation-invariant form
\begin{equation}
\phi(r,\varphi,z)=ve^{i\varphi n_W}f(r),
\label{equ:vortex}
\end{equation}
where $n_W$ is the winding number and $f(r)$
is a function that vanishes at $r=0$ and approaches $1$ at infinity.
In three dimensions, the vortices are one-dimensional vortex lines
(see Fig.~\ref{fig:string}),
and in the cosmological context, they are also known as {\it cosmic strings}.

We have seen that the time zones correspond to a
spontaneous breakdown of a U(1) symmetry, 
and there are also vortices in this analogy, namely the North and
South Poles.
Around each pole, the time zone changes by a full 24 hours
around any closed path that encircles a pole. The time zone
is not well defined at the pole itself, and the symmetry is therefore
restored there.

For a three-component real scalar field, the
vacuum manifold is a sphere. In that case, it is possible that the field
values on a closed surface $S$ in space wrap around the sphere, analogously 
to how the field values wrapped around the circle on the curve $C$ if the 
winding number was non-zero.
Then the same argument shows that there must be a point
inside the surface at which the field leaves the vacuum manifold.
This is a {\it monopole}. A typical monopole
has the spherically symmetric ``hedgehog'' form,
\begin{equation}
\Phi^a(\vec{x})=vf(|\vec{x}|)\frac{x^a}{|\vec{x}|},
\label{equ:hedgehog}
\end{equation}
where $v$ is a constant. The function $f(r)$ is again a function that
vanishes at $r=0$ and approaches unity at infinity, 
which means that the field is on the vacuum manifold far away from the origin.
In $\Phi^a$ the index $a$ indicates the direction in the internal space, 
whereas in $x^a$ it indicates the direction in coordinate space. 
In this sense, $\Phi$ and $x$ are parallel.

With a four-component scalar field, there is a further possibility 
that the field configuration in the whole three-dimensional
coordinate space wraps around the vacuum manifold. That is known
as a {\it texture}~\acite{Davis:1986nr}. 
It is easy to see that a texture is not a stable object, because the energy 
of any scalar field configuration is generally of the form
\begin{equation}
E=\int d^3x \left[ ({\bf \nabla}\phi)^2+V(\phi)\right].
\label{equ:scalarenergy}
\end{equation}
In a texture, the field is everywhere on the vacuum manifold, and the 
latter term vanishes.  The energy of a texture is 
completely due to the spatial variation of the field $\phi$.
If we double the size of the field configuration, the 
first term doubles, and therefore a texture
would shrink 
to a point and disappear in order to minimize its energy.

In fact, only domain walls are truly localized objects in global
theories, because the energies of vortices and monopoles diverge
logarithmically and linearly with the system size, respectively.
This is again due to the gradient term in Eq.~(\ref{equ:scalarenergy}).
The only way to have a configuration with a finite energy is to
have an equal number of vortices and antivortices, or monopoles
and antimonopoles. Even then, a vortex and an antivortex would
have a logarithmic interaction, which binds them in a pair. For a
monopole-antimonopole pair, the interaction is linear, and
therefore they would be confined just like quarks.

Topological defects also exist in gauge field theories, but their
properties are somewhat different. Far from the defect, where the
scalar field is on the vacuum manifold, the gauge field can cancel
the gradient contribution to the energy in Eq.~(\ref{equ:scalarenergy}), 
because the gradient ${\bf\nabla}\phi$ gets replaced by the covariant 
derivative ${\bf D}\phi$ (see Eq.~(\ref{equ:covder})).
In the case of
the Abelian Higgs model, this happens when
\begin{equation}
\vec{A}=-\frac{1}{e}\vec{\nabla}\theta,
\label{equ:gaugemin}
\end{equation}
where $\theta=\arg\phi$ is the phase angle of the scalar field.
Thus, the (cross-sectional) energy of the vortex is finite,
but it still does not vanish
completely, because the energy density
is non-zero near the vortex core, where the
field is away from the vacuum manifold. In any case, this means
that the energy of a defect is localized and finite in gauge field
theories. Their interactions are also much weaker than those of
global defects: Interactions of vortices are exponentially
suppressed at long distances, and monopoles have a Coulomb type
interaction.

It also follows from Eq.~(\ref{equ:gaugemin}) that a gauged vortex
carries magnetic flux~\acite{Nielsen:cs}. 
The phase angle changes by an integer multiple
of $2\pi$ around a
vortex, but this means that the contour integral of the gauge field
around the vortex
must be a multiple of $2\pi/e$. This contour integral is nothing but
the magnetic flux through the contour, 
and therefore the flux is proportional to the
winding number
\begin{equation}
\Phi
=
\oint {\bf dr}\cdot{\bf A}
=-\frac{2\pi}{e}n_W.
\end{equation}
In other words, each vortex carries an integer
multiple of the {\it flux quantum} $\Phi_0=2\pi/e$.

This can be seen by placing a superconductor in an external
magnetic field. As observed earlier, the superconductor repels the
magnetic field, but in many cases, if the field is strong enough,
it will penetrate the superconductor and form Abrikosov flux tubes,
which are exactly what we have called gauged vortices.
Superconductors for which this happens are said to be of Type~II.
In Type~I superconductors, the magnetic field forms one thick
vortex that carries the whole magnetic flux. If $\lambda<e^2$, the
Abelian Higgs model behaves like a Type~I superconductor. Vortices
with multiple windings are stable and two
vortices of equal sign attract each other. In the opposite
case with $\lambda>e^2$, two equal-sign 
vortices repel each other, and the theory resembles a Type~II
superconductor.

The simplest example of a gauge theory with monopoles is the {\it
Georgi-Glashow model}. It consists of a real three-component scalar field
$\Phi^a$, where $a\in\{1,2,3\}$, 
and a three-component gauge field $\vec{A}^a$.
The gauge field makes the theory invariant under local
(i.e. position-dependent) rotations of $\Phi^a$ in the
three-dimensional internal space.
This symmetry
is spontaneously broken by a vacuum expectation value
$\Phi^a\Phi^a=v^2$. However,
it is still possible to do rotations with $\Phi^a$
as an axis without changing it, and therefore there is a residual U(1)
symmetry. We have seen before that a local U(1) symmetry gives rise to
the electromagnetic field, and this happens in this case, too.

In normal electrodynamics, the field lines of the magnetic field
cannot end, which means that there cannot be magnetic
charges. The same
is true in the broken phase of the Georgi-Glashow model as long as
$\Phi^a\ne 0$. However,
$\Phi^a$ vanishes at the centre of the monopole solution
in Eq.~(\ref{equ:hedgehog}), and 
the corresponding solution in the
Georgi-Glashow model does indeed have a non-zero magnetic charge.
This solution, which is known as the {\it
't~Hooft-Polyakov monopole}~\acite{'tHooft:1974qc,Polyakov:ek}, 
is therefore a magnetic monopole. 
The total energy
of a 't~Hooft-Polyakov monopole is finite, but it has a long-range
$1/r$ contribution. Because the monopoles have magnetic charges,
it is easy to understand this as
the magnetic analogue of the Coulomb interaction.

Similar 't~Hooft-Polyakov monopoles exist in essentially 
all theories with a residual U(1) subgroup. In particular, this
includes all GUTs, and therefore they generally predict the existence
of magnetic monopoles. The masses of these monopoles are determined by the
energy scale at which the U(1) group appears.
In the SU(5) GUT, the monopoles would be
extremely heavy, $m_M\approx 10^{17}$~GeV, but for instance in SO(10),
somewhat lighter monopoles with $m_M\approx 10^{13}$~GeV would be
possible (see Section~\ref{sect:particle}). 

\section{Early universe}
\label{sect:universe}

It was observed by Hubble in 1929 that the light of
distant galaxies is redshifted,
and that the amount of
redshift is proportional to the distance. The simplest explanation
to this is that the galaxies are moving away from us and the
redshift is caused by the Doppler effect. Because the velocity is
proportional to the distance from us, this does not imply that we
occupy a special position at the ``centre'' of the universe, but
that the universe is expanding with the same rate everywhere.

In order to study the evolution of the whole universe, we assume
the {\it cosmological principle}, which means that the universe
looks the same at every point and in every direction, which seems to
be a good approximation at very long distances. In technical terms,
we assume that the universe is homogeneous and isotropic.

This assumption, together with the observation that the universe is
spatially flat~\acite{Spergel:2003cb} means that we can describe its expansion by two
equations, the Friedmann equation and the conservation of energy,
\begin{eqnarray}
\left(\frac{\dot a}{a}\right)^2&=&\frac{8\pi G}{3}\rho,
\label{equ:Friedmann}\\
\dot\rho&=&-3\left(\rho+\frac{p}{c^2}\right)\frac{\dot a}{a}.
\label{equ:consE}
\end{eqnarray}
Here $a(t)$ is the {\it scale factor} of the universe, the dot 
indicates a time derivative, $\rho$ is energy density and $p$ is pressure.
The scale factor shows how
the space stretches. The distance between two objects at rest
at time $t_1$ is longer by a factor of $a(t_1)/a(t_0)$ than it
was at time $t_0$. 
As the Hubble parameter $H=\dot{a}/a$ is positive, the universe is
expanding, and distant galaxies would indeed appear to be moving away
from us.

Conservation of energy~(\ref{equ:consE}) shows that as 
the universe expands, the
energy density decreases. Consequently, the temperature of the
universe has been higher in the past. If we know the
equation of state, i.e. the pressure $p$ as a function of $\rho$
and $T$, we can use Eqs.~(\ref{equ:Friedmann}) and
(\ref{equ:consE}) to follow the time evolution of the scale factor
$a$ arbitrarily far back in time. If we assume that the equation
of state does not change dramatically, this analysis shows that
the scale factor started from zero about 13.7 billion 
years~\acite{Spergel:2003cb} 
ago.
We should not take this result too literally, because at that time,
the energy density would have been infinite and we have no reason
to believe that our assumptions were valid.

In any case, we can quite confidently
extrapolate the evolution of the scale factor almost all the way back
to this initial singularity, which we can choose to be at $t=0$.
By observing the cosmic microwave
background (CMB) radiation, we can actually {\it see} all the way to
$t\approx 380000$~years~\acite{Spergel:2003cb}, when the universe became
transparent. Although we
cannot see beyond this ``surface of last scattering'', we can use
theoretical calculations to find out what happened before that.

Atomic and nuclear physics, and particle physics that has been
tested and confirmed in accelerator experiments, can take us all the
way back to $t\approx 10^{-10}$~s, which corresponds to the
temperature of $100$~GeV.
To reach ever earlier times, we have to make assumptions about physics
that is out of the reach of any experiments. On the other hand, we can use
the universe as the ultimate accelerator experiment, and calculate
what kind of observational predictions follow from a given
theory. This gives us a way of reaching much higher energies than
particle accelerators ever can, albeit with the price that we cannot
repeat the experiment.

As we will see in Section~\ref{sect:transitions}, symmetries that are
broken at zero temperature are usually unbroken at high
temperatures. If the temperature of the universe was initially high enough,
the universe
would have undergone a phase transition from the unbroken to the broken phase
as it expanded
and cooled down. We therefore expect that there were a number of symmetry
breaking phase transitions in the early universe, and they may have
led to defect formation.

On the other hand, this hot Big Bang scenario
cannot have been valid all the way to the initial singularity, and
perhaps the most compelling reason is the horizon 
problem~\acite{Guth:1980zm}. 
Different directions on the sky look very
similar, and if we actually measure the temperature of the cosmic
microwave background, it is uniform up to one part in 100~000 in different
directions. Yet, this light was emitted only $380000$ years after the
initial singularity, during which time any information could only have
travelled a distance of about 1 degree on the sky. Therefore, there is
no physical mechanism that could have made different directions reach
thermal equilibrium with each other. There are also other problems with 
the Big Bang scenario, such as the absence of magnetic
monopoles, to which we will return later.

\begin{figure}[t]
\center
\epsfig{file=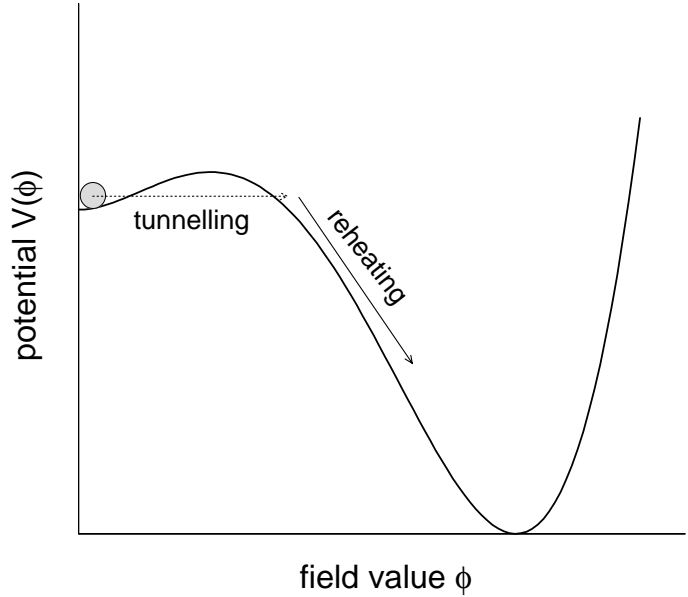,width=9cm}
\caption{
\label{fig:metastable}
The shape of the potential in Guth's original inflationary model. The
field, represented by the gray ball, is initially in the local
minimum, and the non-zero potential energy makes the universe expand
with a constantly accelerating rate. Finally, the field tunnels
through the potential barrier, and inflation ends. The non-zero
potential energy is released during reheating, and the universe
reaches thermal equilibrium.
}
\end{figure}

In order to solve these problems, Alan Guth proposed the theory of
inflation~\acite{Guth:1980zm} 
(see also Refs.~\acite{lindebook,Copeland:gs}), 
whose basic idea is that the universe
was dominated by vacuum energy at some early stage.
This can be achieved with a real scalar field, which is known as the
{\it inflaton} and which has a
potential with the shape shown in Fig.~\ref{fig:metastable}. The
energy density $\rho$ is mostly due to the kinetic energy $\dot\phi^2/2$ and
the potential energy $V(\phi)$.
If the field is initially placed at the local minimum at
$\phi=\phi_0$, the time derivative $\dot\phi$ vanishes, and
only the vacuum contribution $V(\phi_0)$ remains.
Because
$\phi$ does not change, the energy density $\rho$ is constant and
Eq.~(\ref{equ:Friedmann}) implies that
the scale factor $a(t)$ grows exponentially.

Inflation not only explains why the universe is so homogeneous, but
also where all the structure we see in it today came from. The exponential
expansion amplifies quantum fluctuations, and they act as seeds for
structure formation and can be seen as fluctuations in the CMB.
Recent measurements of these fluctuations~\acite{Spergel:2003cb} 
agree remarkably
well with the predictions.

In order to explain the observed homogeneity over the whole sky,
inflation must have expanded the universe by at least a factor of
around $10^{20}$, but after that, it must have come to an end in
order to match with the ordinary Big Bang cosmology. 
Because energy is conserved, the energy density that was driving inflation,
was released as radiation during this
process. The dynamics of this {\it reheating} depends on the details
of the inflationary model, but eventually the universe reached thermal
equilibrium at the {\it reheat temperature} $T_{\rm rh}$. Its value is
not known, but the asymmetry between matter and antimatter requires it
to be at least $100$~GeV.

In Guth's original inflationary model did not describe reheating in a 
satisfactory way, but there
are a vast number of other inflationary models that have been proposed
to address this and other unattractive
features of the original proposal. Most of them are based on the
assumption of slow roll. In an expanding universe, the equation of
motion (\ref{equ:realeom}) is modified to
\begin{equation}
\ddot\phi+3H\dot\phi-\vec\nabla^2\phi=-V'(\phi),
\label{equ:realeom_exp}
\end{equation}
where $H=\dot a/a$ is the Hubble parameter. The extra term in this
equation behaves as a friction term and slows down the motion of
the field in an expanding universe. Just like an object falling in
the atmosphere reaches a terminal velocity due to the air drag,
the scalar field
reaches a slow roll solution
\begin{equation}
\dot\phi=-\frac{V'(\phi)}{3H}.
\end{equation}
If the slope of the potential is small compared with the Hubble rate,
the value of the potential changes only very slowly.
The energy density $\rho$ is, again, almost constant, and
we find an exponentially growing, inflationary, solution.

\begin{figure}[t]
\center
\epsfig{file=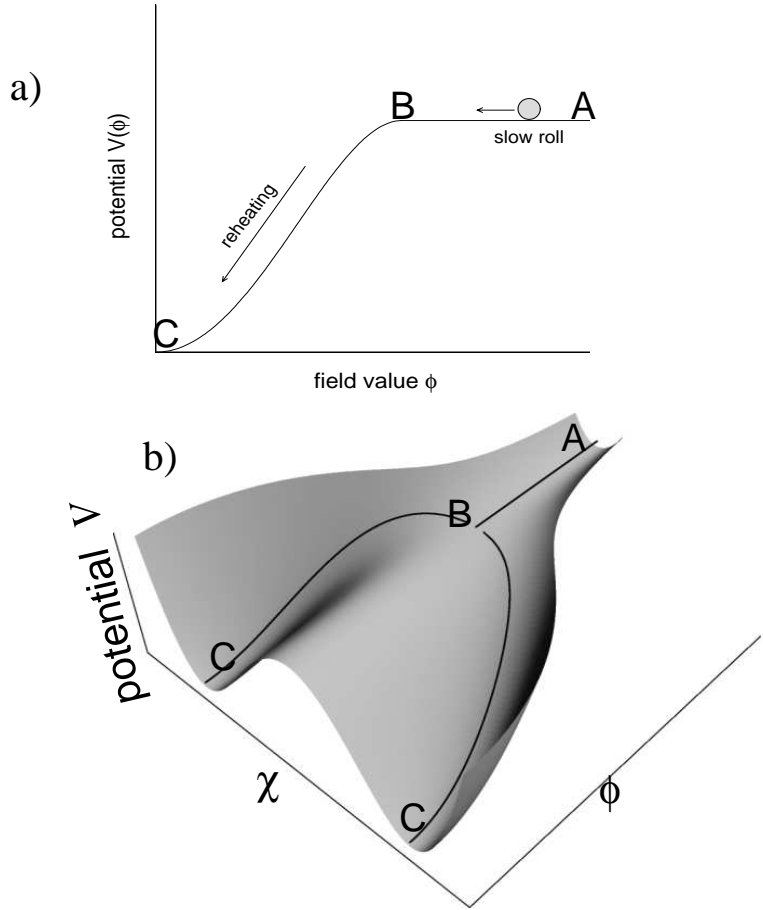,width=10cm}
\caption{
a) A typical potential for slow roll inflation. During inflation (A), the
field is in a part of the potential that is almost flat, and therefore
it is moving very slowly. This gives rise to almost exponential
expansion. Inflation ends when the potential becomes steeper (B). The
universe reheats, and the inflation field ends up in the minimum of the
potential (C).
b) The potential for hybrid inflation. At any time, the waterfall
field \protect{$\chi$} is at its minimum, and therefore the potential is
effectively of the type shown in (a). At the end of inflation, the
\protect{$\chi \leftrightarrow -\chi$} symmetry is broken spontaneously at a
phase transition.}
\label{fig:hybrid}
\end{figure}

In this setup, inflation can end smoothly if the potential starts to
fall steeply at some value of $\phi$, as shown in
Fig.~\ref{fig:hybrid}. Perhaps the most elegant way of achieving that is by
adding another scalar field $\chi$ with the potential
\begin{equation}
V(\phi,\chi)=\frac{1}{2}m^2\phi^2+\frac{1}{2}g^2\phi^2\chi^2+
\frac{1}{4}\kappa\left(\chi^2-v^2\right)^2.
\label{equ:hybrid}
\end{equation}
The shape of this potential is shown in Fig.~\ref{fig:hybrid}, and it is
symmetric under $\chi\rightarrow -\chi$.
This model in known as {\it hybrid inflation}~\acite{Linde:1993cn}.

We can see
that if $\phi$ has initially a large value and $\chi$ is zero, the
fields roll down slowly along the valley towards the origin and the
universe inflates. However, when $\phi$ reaches $\phi_0=\kappa^{1/2}v/g$, 
the
``waterfall'' field $\chi$ has to choose between the positive and
negative branches, and this breaks the symmetry spontaneously. In
either branch, the potential is much steeper, and therefore the 
slow roll assumption fails and inflation ends. The interactions between the
inflaton and other fields reheat the universe as the inflaton rolls 
quickly down the hill to the minimum.

There are many variants of this hybrid inflationary model.
One attractive feature of the models in this class
is that potentials of the form of Eq.~(\ref{equ:hybrid}) arise
relatively naturally in many particle physics models. As we have seen,
inflation is ended by a symmetry breaking phase transition in these
models, and it is therefore important to understand what happens at
this transition and, in particular, whether topological defects are
formed.

\section{Thermal phase transitions}
\label{sect:transitions}

At zero temperature, the equilibrium states of the system
correspond to the minima of the potential.
However, it is very well known
from many everyday systems that a symmetry that is spontaneously
broken at low temperatures can be restored at higher temperatures.
One example of this is water: The crystal structure of ice breaks
the rotation invariance, but when the temperature is above
0\degrees C, ice melts and the rotation invariance is restored.

In field theories, one 
can see this by defining the effective potential $V_{\rm
eff}(\phi)$, which is essentially the free energy of a
system in which the spatial average of the scalar field
$\phi(\vec{x})$ is constrained to be
$\phi$. Because the free energy is minimized in
thermal systems, the equilibrium state corresponds to the minimum
of the effective potential.

\begin{figure}[t]
\center
\begin{tabular}{ll}
a)&b)\\
\epsfig{file=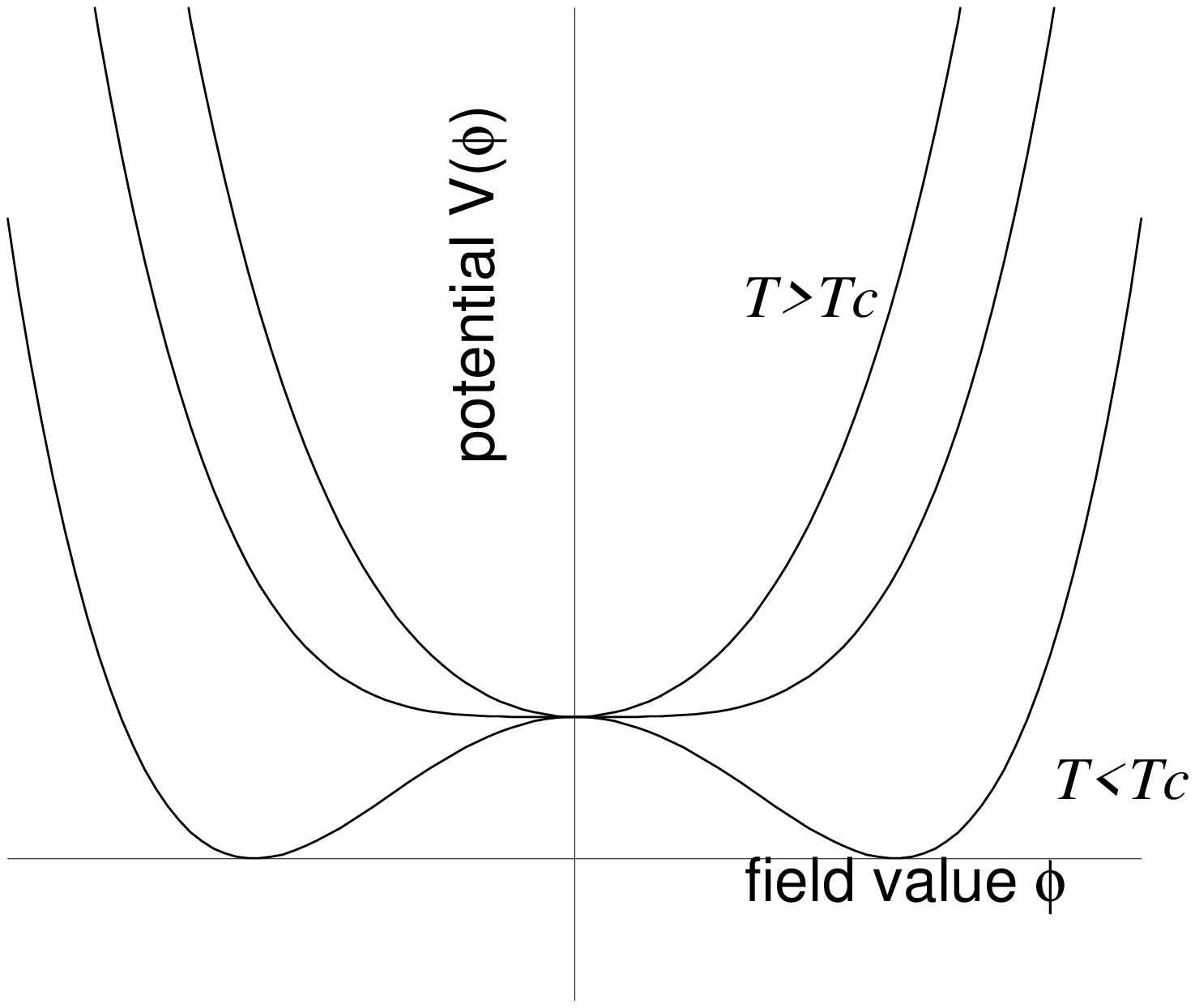,width=6cm}
&
\epsfig{file=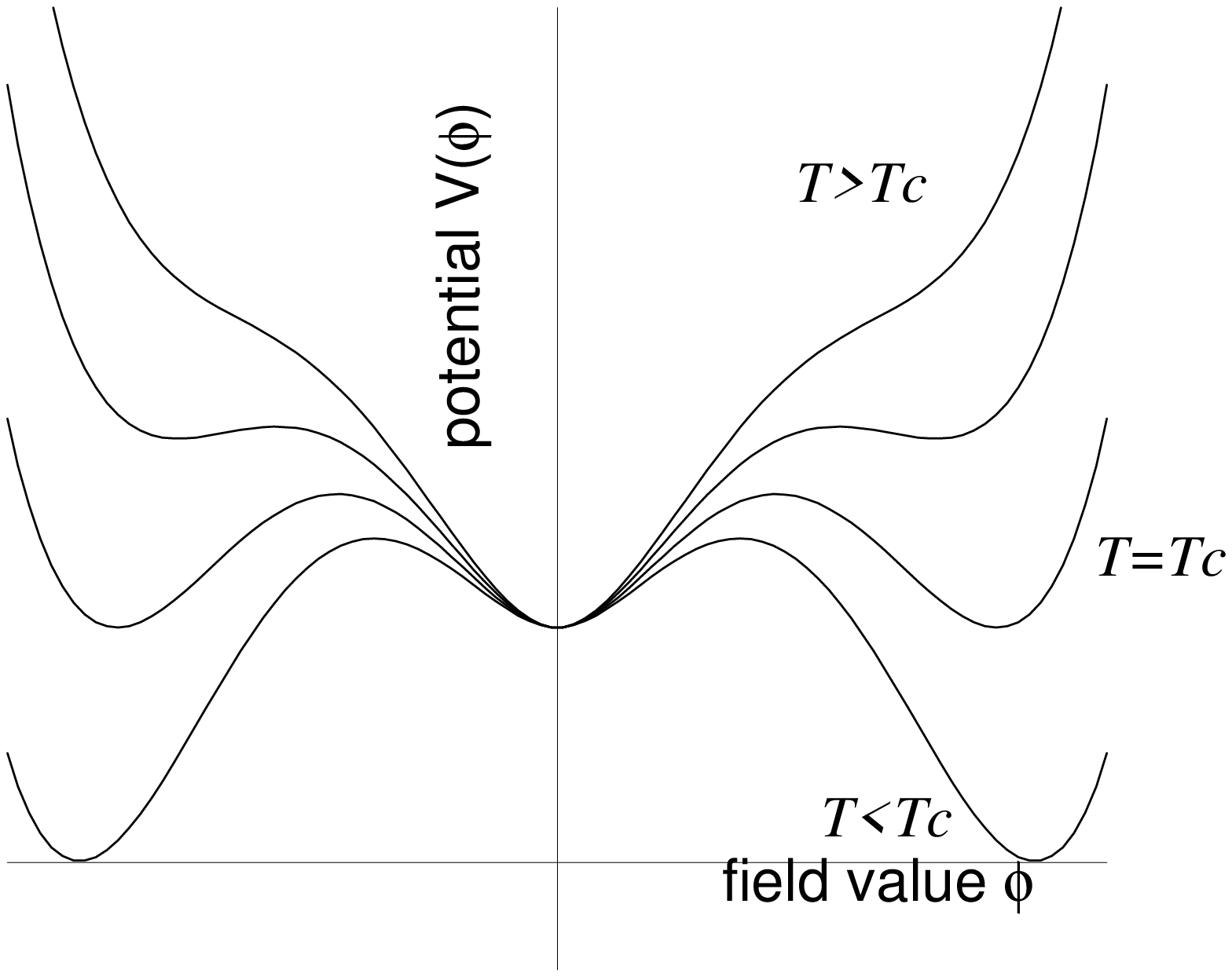,width=6cm}
\end{tabular}
\caption{
\label{fig:symmrest}
a) The effective potential 
in a continuous phase transition. Above the critical temperature
$T_c$, the origin $\phi=0$ is the only minimum, and the symmetry is
restored. Two degerate symmetry-breaking minima appear at $T=T_c$ and
the origin becomes unstable.
b) The effective potential in a first-order phase transition. 
At very high temperatures, the origin $\phi=0$, is the only minimum,
but when the temperature goes down, two metastable symmetry-breaking
minima appear. At the critical temperature $T_c$, these minima are
degenerate with the symmetric one. Below $T_c$, the symmetric minimum
becomes metastable and eventually decays through bubble nucleation.
}
\end{figure}

The effective potential  is typically 
calculated using perturbation
theory~\acite{Jackiw:cv}, 
and the leading contribution comes from one-loop Feynman
diagrams. In the scalar theory, the main effect is to give a
temperature-dependent quadratic contribution to the potential, $\Delta
V(\phi)\propto \lambda T^2\phi^2$~\acite{Kirzhnits:ut,Dolan:qd}.
As shown in Fig.~\ref{fig:symmrest}a, this restores the symmetry when
the temperature is
above the
critical temperature $T_c\approx v$.

The minimum of the effective potential gives the mean value
$\langle\phi\rangle$, but the field fluctuates
around this value. Therefore, even if the symmetry is restored and
the mean value $\langle\phi\rangle$ vanishes, the actual value of
the field at any given point is typically non-zero. In fact, the
field only vanishes in a set of measure zero at any given time.

To leading approximation, the thermal fluctuations have a Gaussian
distribution and therefore their properties are specified by the
two-point function, which typically has an exponentially decaying
form. The decay rate is characterized by the
{\it correlation length} $\xi$. In
practice, it means that fluctuations at two points separated by a 
distance greater than $\xi$ are independent.

When the transition point is approached, the correlation length
diverges with a critical exponent $\nu$,
\begin{equation}
\xi(T)\sim \left(\frac{T-T_c}{T_c}\right)^{-\nu},
\label{equ:nudef}
\end{equation}
which depends on the universality class of the system. At the same
time, the dynamics of the system becomes slower, and this critical
slowing down can be expressed in terms of the relaxation time
$\tau(T)$, which diverges with a (different) critical exponent
$\mu$,
\begin{equation}
\tau(T)\sim \left(\frac{T-T_c}{T_c}\right)^{-\mu}.
\label{equ:mudef}
\end{equation}
For a real scalar field $\nu\approx 0.630$~\acite{Hasenbusch:1998gh}, 
and for a complex
field $\nu\approx 0.672$~\acite{Hasenbusch:1999cc}. 
In most systems, $\mu=1$.

Unfortunately, it is often not possible to use the simple
description in terms of the effective potential in gauge field
theories. Thermal fluctuations can easily change the field
configuration to any of its gauge transforms, because that does not
cost any energy. This means that if we average over these
fluctuations, the mean value $\langle\phi\rangle$ is always zero. 
This simple observation means
that a local gauge symmetry cannot strictly speaking be broken
spontaneously~\acite{Elitzur:im}, and that one should only consider manifestly
gauge-invariant quantities.

Nevertheless, it is possible to calculate the effective potential,
if one first fixes the gauge. That means imposing a constraint,
which breaks the gauge invariance, but does not change any
gauge-invariant quantity. Then, one can calculate the effective
potential and use it to study the phase transition. This may give
non-zero values to non-gauge-invariant quantities such as
$\langle\phi\rangle$, 
but one should keep in mind that this is just a
non-physical consequence of the gauge fixing.

Even after the gauge fixing, the calculation of the effective
potential is difficult in gauge field theories, and the reason is
that the gauge bosons are massless in the symmetric phase. 
If the field in the
one-loop Feynman diagram that contributes to the effective
potential is massless, the diagram diverges and the calculation breaks down. 
This infrared problem
can be partly solved by using a trick called resummation, but even
that only works when the scalar coupling $\lambda$ is weak compared to
the gauge coupling $g$~\acite{Kirzhnits:1976ts}. 
In that case, there is an
extra cubic contribution $g^3T\phi^3$ to the effective potential.
There is, therefore, a range of temperatures in which the
potential has two minima, one at zero and one at a non-zero 
value (see Fig.~\ref{fig:symmrest}b).
Furthermore, the values of a gauge-invariant quantities such
as $\langle|\phi|^2\rangle$ are different in these two minima, and therefore
they actually correspond to physically different states.

When the minimum that corresponds to non-zero
$\langle\phi\rangle$ becomes the global minimum, we have exactly the same
situation as in Guth's original inflationary model. The symmetric
vacuum becomes metastable, and bubbles of the true, broken vacuum
start to nucleate, in perfect analogy with bubbles of vapour in boiling 
water. A transition like this is known as a {\it
first-order phase transition}.
This does not always lead to inflation, though, because in many cases
the energy of the thermal fluctuations dominates over the potential
energy.  

On the other hand, when the scalar coupling is strong relative to the
gauge coupling, the perturbative
calculation of the effective potential is unreliable. Generally,
one will then have to use numerical lattice Monte Carlo
simulations. It is known that for the Abelian Higgs
model there is a phase transition, and the two phases can be
characterized by whether the magnetic field has long-range
correlations or, in other words, whether there is Meissner effect
or not.

These correlations can be seen in the thermal fluctuations of the
magnetic field, which are actually nothing but thermal radiation,
which we will approximate by blackbody radiation.%
\footnote{
Because we are
interested in long wavelengths, the fluctuations are described by the 
classical Rayleigh-Jeans spectrum (\ref{equ:RJspectrum}). 
This spectrum cannot be valid at very short wavelengths
where $k\gsim T$,  because it would lead to the famous ultraviolet
catastrophe. This was the reason why Planck postulated that the energy 
of radiation is quantized, which eventually led to the development of 
quantum mechanics. However, because we are considering macroscopic effects,
the Rayleigh-Jeans spectrum is sufficient.}
We can think of the fluctuations as a
superposition of plane waves with different wave numbers $\vec{k}$,
each of which has a Gaussian probability distribution with variance
$G(k)$. In the symmetric phase, this is simply equal to the energy of
the degree of freedom, and because of classical equipartition of
energy, we have
\begin{equation}
G(k)=T.
\label{equ:RJspectrum}
\end{equation}
In the broken phase, the Meissner effect starts to suppress
long-wavelength fluctuations, and 
we find
\begin{equation}
G(k)=\frac{T}{1+m_\gamma^2/k^2},
\label{equ:Meissner}
\end{equation}
where $m_\gamma$ is the inverse of the photon correlation length, which
is also known as the penetration depth in superconductivity. Near the
transition in the broken phase, 
the value of $m_\gamma$ is approximately given by
\begin{equation}
m_\gamma^2\approx e^2(T_c^2-T^2).
\end{equation}

 In non-Abelian theories such as the electroweak theory,
the Georgi-Glashow model or GUTs, the transition actually
disappears and becomes a smooth crossover when the scalar
self-interactions dominate over the gauge coupling~\acite{Hart:1996ac}. 
This may sound odd, because one might have thought
about using the value of  $\langle\phi\rangle$ to distinguish
between the two phases, but this quantity is not gauge-invariant
and does not therefore have a physical interpretation. In fact,
because the gauge symmetry cannot be broken, there is no
qualitative difference between the two phases, and in this sense
the disappearance of the transition should not be surprising. After all,
exactly the same happens to the water-vapour transition at a high
enough pressure.

\section{Global symmetries -- the Kibble-Zurek mechanism}
\label{sect:kibble}

Let us first consider what happens in theories with global
symmetries if a symmetry that was initially present gets
spontaneously broken. The equilibrium state
of the system after the transition should in principle correspond
to a constant, non-zero value of $\phi$ with some small thermal fluctuations
around it. However, the choice of $\phi$ is not unique, because it
could equally well be any point on the vacuum manifold. 
It was pointed out by Tom Kibble
in 1976 that this leads to defect formation~\acite{Kibble:1976sj}.

If we think of two points that are far enough from
each other, there is no reason to expect that they would make the
same choice. In the cosmological context, an obvious constraint is
that any two points 
must make this choice independently of each other if they are so
far from each other that even a light signal would not have had time to
travel from
one to the other. 
However, it is equally obvious that
this only gives the upper limit for the distance. For the
moment, let us just say that there is a critical length scale
$\hat\xi$, and at longer distances the choice of the vacuum is
uncorrelated.

In this case, the system cannot reach the true equilibrium state,
because that  would require making the same choice at infinite distances.
Instead, the best it can do to minimize the energy
is to make the field more or less constant
at distances less than $\hat\xi$. To get a picture of this, we can
think that it forms domains of radius $\hat\xi$, inside each of
which the field is constant and between which it is random.

\begin{figure}[t]
\center
\epsfig{file=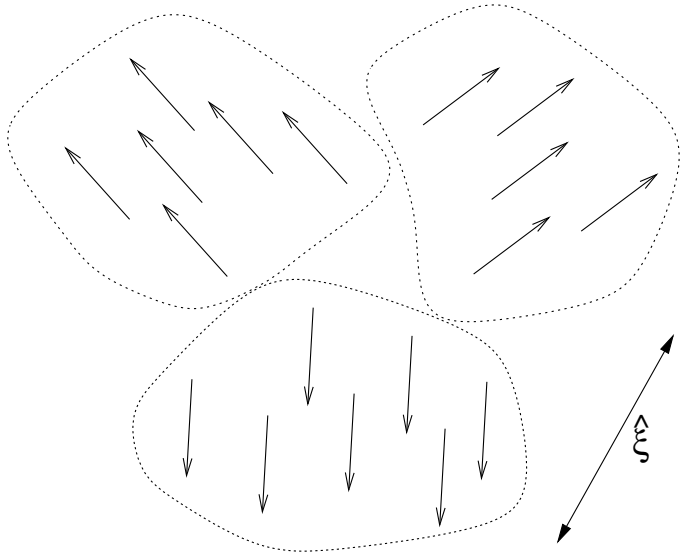,width=9cm}
\caption{
\label{fig:domains}
Vortex formation by the Kibble mechanism. Because the system has only
a finite time available, it can only be ordered at distances less than
a critical value $\hat\xi$. Therefore, it forms domains of these size,
inside each of which the scalar field, depicted by the small arrows,
is aligned. The system tries to interpolate the field between the
domains, but it can only do it if the field vanishes somewhere between
the domains. The vortex appears where the field vanishes.
}
\end{figure}

Of course, there would not really be well-defined domains, because
the field would vary smoothly from one domain to another. 
However, even that is not possible everywhere. Imagine, for instance, that a
U(1) symmetry is broken in two dimensions. Fig.~\ref{fig:domains} shows a
possible configuration of three neighbouring domains, and one can
interpolate the field from 1 to 2, from 2 to 3 and from 3 to 1,
but if one does that, one finds that there is a vortex at the
point where these domains meet. This means that there is roughly 
one vortex per domain, and the number density
of vortices formed in the transition is
\begin{equation}
n\sim 1/\hat\xi^2. \label{equ:stringKibble}
\end{equation}
The same estimate applies to strings in three dimensions, except
that that number density should then be interpreted as the number
density per cross-sectional area. In other words, the transition
produces a string network, and the number of strings that cross any
two-dimensional plane is given by Eq.~(\ref{equ:stringKibble}).

It is easy to generalize this argument to monopoles, and that
yields a number density per unit volume of
\begin{equation}
n\sim 1/\hat\xi^3. \label{equ:monopoleKibble}
\end{equation}
Because global monopoles have very strong interactions due to
their linear interaction potential, this expression is only valid
immediately after the transition. Later on, the interaction starts
to pull monopoles and antimonopoles together, and whenever they
meet they annihilate. The number density of monopoles would
therefore drop very quickly from its initial value in
Eq.~(\ref{equ:monopoleKibble}). To a somewhat lesser extent, the
same is also true for global vortices, which have a logarithmic
interaction potential.

The predictions in Eqs.~(\ref{equ:stringKibble}) and
(\ref{equ:monopoleKibble}) are not very useful unless we know how to
calculate $\hat\xi$. It was argued by Zurek in 1985~\acite{Zurek:qw} 
that it depends
mostly on the critical exponents $\mu$ and $\nu$ defined in 
Eqs.~(\ref{equ:nudef})
and (\ref{equ:mudef}). 
For simplicity, we assume that the temperature is decreasing
linearly,
\begin{equation}
T(t)=\left(1-\frac{t}{\tau_Q}\right)T_c.
\end{equation}
The quantity $\tau_Q$ characterizes the cooling rate, and is known as
the {\it quench timescale}.
When the temperature decreases towards $T_c$, the correlation
length $\xi$ grows as $\xi(t)\sim (|t|/\tau_Q)^{-\nu}$, 
but at the same time the
dynamics of the system becomes slower, 
$\tau(t)\sim (|t|/\tau_Q)^{-\mu}$. Eventually, at some time $\hat t$,
the relaxation time $\tau$ becomes equal to the time that is left
before reaching $T_c$, i.e.,
$\tau(\hat t)=|\hat t|$. After this time, 
the system cannot adjust to the change of
the temperature fast enough, and falls out of equilibrium.
This freeze-out time depends on $\tau_Q$ as
$
\hat t \propto -\tau_Q^{\mu/(1+\mu)}.
$
The correlation length cannot grow significantly after this time, and
therefore we can, to a good approximation, say that it freezes to 
whatever its value was
at time $\hat t$, and that determines the freeze-out scale
$\hat\xi$~\acite{Zurek:qw},
\begin{equation}
\hat\xi=\xi(\hat t)\propto \tau_Q^{\nu/(1+\mu)}.
\label{equ:Zurek}
\end{equation}

A similar, but much simpler, picture can be used for first-order transitions
that take place through bubble nucleation.
In that case, the domains
would be replaced by bubbles, and the scale $\hat\xi$ would be
the typical bubble radius at the time when the bubbles coalesce.

\begin{figure}[t]
\center
\epsfig{file=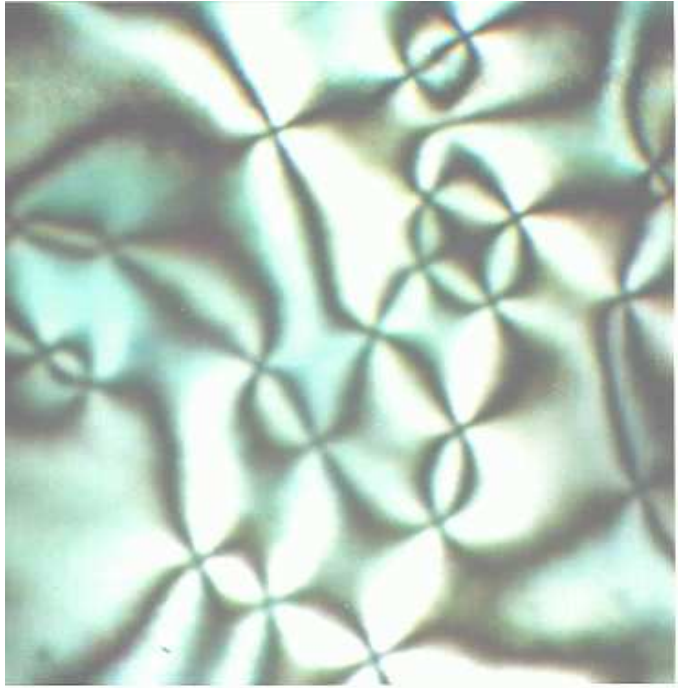,width=9cm}
\caption{
\label{fig:lcfig}
Snapshot of a liquid crystal after the phase transition. 
Vortices are the places where the dark regions
form a cross.
(From Digal et al.~\cite{Digal:1998ak})}
\end{figure}

The predictions  Eq.~(\ref{equ:stringKibble}) and (\ref{equ:Zurek})
have been  tested
experimentally in a variety of systems, ranging from superfluid
$^4$He~\acite{Hendry,Dodd} 
and $^3$He~\acite{Ruutu:1996qz,Bauerle} 
and liquid crystals
(see Fig.~\ref{fig:lcfig})~\acite{TurokNature,Bowick:1994rz,Digal:1998ak} 
to nonlinear optical systems~\acite{Ducci1999} and
convective fluids~\acite{Casado2001}. 
The results agree mostly with the predictions,
although systematic tests have not been possible in many cases,
because it is typically quite difficult to vary the quench time scale $\tau_Q$
and to determine the critical exponents $\mu$ and $\nu$.

Therefore, it is perhaps better to concentrate on
how the defects and antidefects are
distributed in space~\acite{Digal:1998ak}, 
because that is generally insensitive to details
like these. Let us, for simplicity, discuss a
two-dimensional system with vortices and consider the winding number
$N_W(R)$ around a circle of radius $R$. On average, this is of course
zero, $\langle N_W(R)\rangle=0$, 
because there are as many vortices as antivortices. However, 
the variance $\sigma_W(R)\equiv\langle N_W(R)^2\rangle$ is generally
non-zero, and its behaviour as a function of $R$ reflects the spatial
distribution of vortices.

If we assume that the locations of the vortices are more or less
uniformly distributed, the number of vortices plus antivortices inside
a circle of radius $R$ is $N_{\rm vort}=n\pi R^2$, where $n$ is the number
density. 

If each of the $N_{\rm vort}$ vortices were just as
likely to be positive as negative, $N_W(R)$ would be just a result of
a random walk of $N_{\rm vort}$ steps of unit length. The variance of
that is well known to be
\begin{equation}
\sigma_W(R)=N_{\rm vort}\propto R^2.\qquad\text{(Random)}
\end{equation}
However, if the vortices were produced by the Kibble mechanism, we
would expect something different. The arc of the circle would cross
roughly $N_{\rm dom}\approx R/\hat\xi$ 
correlated domains. Each time we move from one domain to another, the
phase angle changes by an essentially random amount, and around the
whole circle, there are $N_{\rm dom}$ of these changes. Therefore, we
have a random walk of $N_{\rm dom}$ steps, and consequently,
\begin{equation}
\sigma_W(R)
\propto N_{\rm dom}
\propto R.~\text{(Kibble)}
\end{equation}
More generally, we can follow Ref.~\acite{Digal:1998ak} and
characterize the distribution of the vortices
by writing
\begin{equation}
\sigma_W(R)\propto R^{4\nu_\sigma}.
\end{equation}
The above arguments show that a random distribution would give
$\nu_\sigma=1/2$, whereas the Kibble mechanism would predict $\nu_\sigma=1/4$.
These predictions do not depend on any quantities that would have to
be measured or calculated separately such as the critical exponents.

The exponent $\nu_\sigma$ was measured experimentally for a phase transition
in liquid crystals by Digal et al.~\acite{Digal:1998ak}, who found
$\nu_\sigma=0.26\pm0.11$, in agreement with what the Kibble mechanism predicts.

\section{Gauge symmetries -- trapping of thermal fluctuations}
\label{sect:trapping}

If the symmetry broken in a phase transition is a local gauge symmetry,
it is not enough to consider the behaviour of
the scalar field~\acite{Hindmarsh:2000kd}. 
The gauge field has a dynamics of its own, and as
Eq.~(\ref{equ:gaugemin}) 
shows in the case of the Abelian Higgs model, the Higgs
field does not even try to be aligned if $\vec{A}$ is non-zero.

Let us start by discussing the Abelian Higgs model. If we imagine
having initially a uniform magnetic field
$\vec{B}=\vec{B}_0$
in the system, we know very well from superconductor experiments
what happens when we cool it into
the broken (superconducting) phase~\acite{superconductors}: 
The magnetic field forms a lattice
of vortices. Because all the flux must go into the vortices, the
number density would be $
n=(e/2\pi)|\vec{B}_0|.$
This happens no matter how slowly we cool the
system, and therefore the formation of these defects 
is clearly not described by the Kibble
mechanism.

Imagine now that the magnetic field is not uniform, but that there is
a magnetic plane wave of some wavelength $\lambda$.
In the absence of
magnetic monopoles, the magnetic field lines cannot end, and the
only way in which a field line loop can disappear is therefore by shrinking
to a point. However, that cannot happen in less time than what it takes for
light to travel from one end of the loop to the other. This
time is simply the wavelength in units of the speed of light, 
and therefore the lifetime of a
fluctuation cannot be shorter than the wavelength. 
It is actually more
convenient to use the wave number $k=1/\lambda$ than the
wavelength $\lambda$. 
If we define the decay rate $\gamma(k)$ as the inverse of the time it takes
for the amplitude of a fluctuation with wavenumber $k$ to halve,
this result gives a causality bound $\gamma(k)<k$.

We can also calculate $\gamma(k)$ more precisely.
The behaviour of the magnetic field is determined by Maxwell's
equations, and in vacuum it would simply oscillate with frequency $k$.
However, in any medium, such as the hot elementary
particle plasma that filled the early universe, there are electric
currents. If they obey Ohm's law with conductivity $\sigma$,
long-wavelength modes ($k\ll\sigma$) have a decay rate
\begin{equation}
\gamma(k)=\frac{k^2}{\sigma},
\end{equation}
which is much slower than the causality bound.

\begin{figure}
\center
\begin{tabular}{ll}
a)& \cr
& \epsfig{file=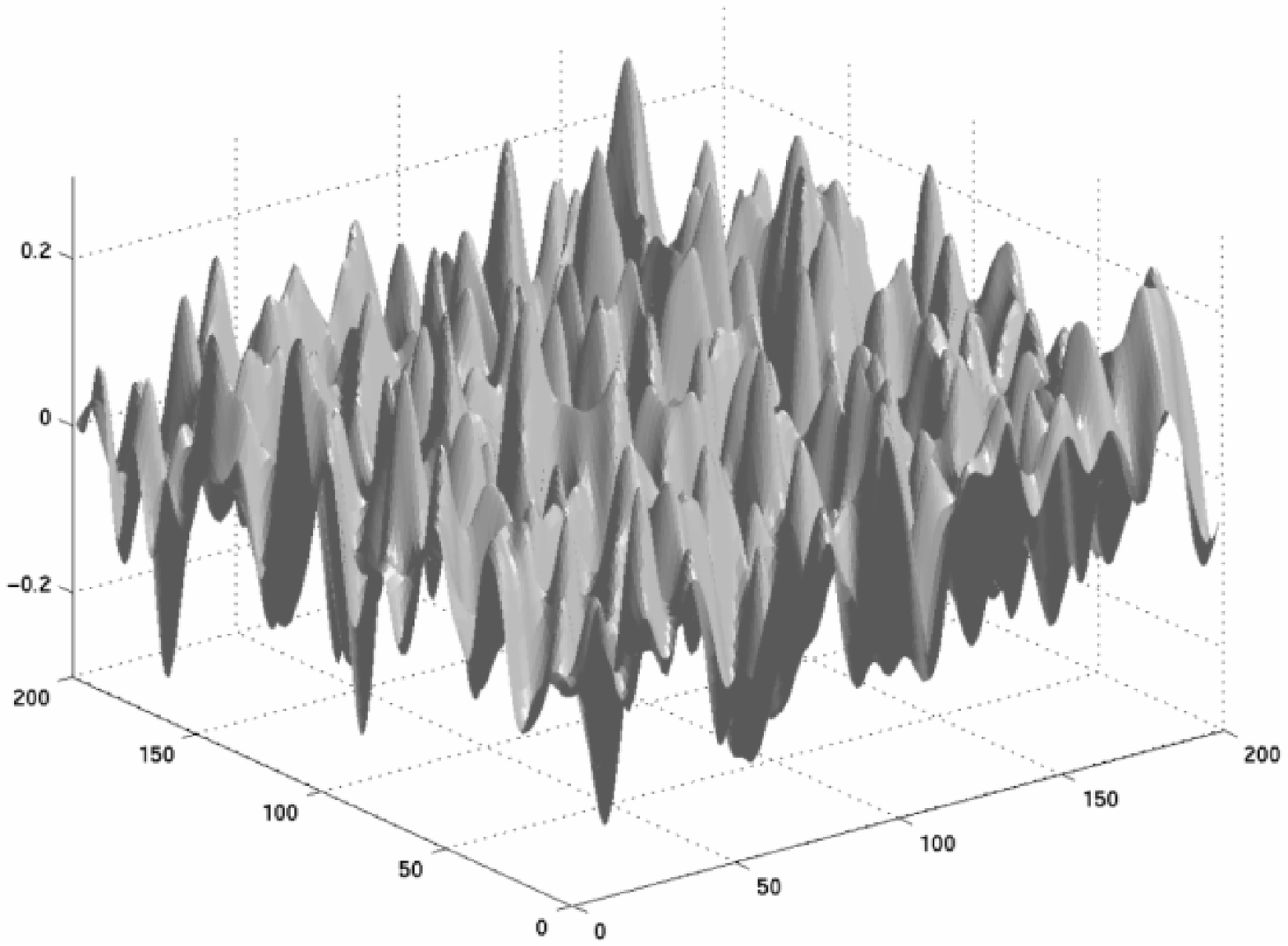,width=8cm}\cr
b)&\cr
& \epsfig{file=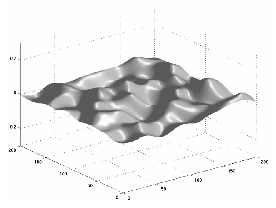,width=8cm}\cr
c)&\cr
& \epsfig{file=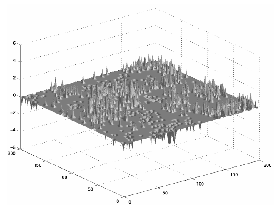,width=8cm}
\end{tabular}
\caption{
\label{fig:trapping}
Vortex formation by flux trapping in the Abelian Higgs model.
a) In the
symmetric phase, there are Gaussian thermal fluctuations of all
wavelengths.
b) During a rapid quench, longest wavelengths $\lambda\gsim 1/k_c$ 
are too slow to react
and freeze out. Shorter wavelengths disappear, leading to a smooth
magnetic field, which varies over the distance scale $1/k_c$.
c) The broken phase is a superconductor, and the frozen-out magnetic
field becomes trapped into vortices.
}
\end{figure}

If we start with the magnetic plane wave and cool the system
into the superconducting phase very rapidly, the wave does not have
time to decay. Because there can
be no magnetic field in the broken phase, it forms vortices
instead, as illustrated in Fig.~\ref{fig:trapping}. 
The vortices point in the same direction as the initial
magnetic field, and their number density is higher where the field was
stronger. In fact, if we looked at the system with a low enough
resolution so that we could not see individual vortices but could
still measure the average magnetic field arising from the flux quanta
carried by the vortices, we would still
see the initial magnetic field.
On the other hand, if the cooling is slow enough, then the initial wave decays
and no vortices are formed.

So, we have seen that a non-zero initial magnetic field leads to
vortex formation, but where could that initial magnetic field come
from? The answer is thermal fluctuations. We already saw that 
arbitrarily long wavelengths are present in the
Rayleigh-Jeans spectrum (\ref{equ:RJspectrum}) of thermal radiation.
Let us consider a single mode with wave number $k$, 
i.e. a single plane-wave component of
the thermal spectrum. If the corresponding 
decay rate $\gamma(k)$ 
is slower than what it would have to be in order for the mode 
to stay in equilibrium, the mode
freezes out and keeps its initial equilibrium amplitude, forming
vortices as we saw above.

If we know how the temperature changes,
we know from Eq.~(\ref{equ:Meissner}),
what the equilibrium amplitude of any given mode
is and therefore how quickly the mode would have to to decay
in order to adjust to this change. 
It is obvious that there are always
going to be some very long wavelengths that are unable to do it. 
Simply from causality, we obtain the lower bound
\begin{equation}
k_c\gsim \left(\frac{e^2T_c^2}{\tau_Q}\right)^{1/3},
\end{equation}
and if Ohm's law is valid, we have
\begin{equation}
k_c\approx \left(\frac{e^2T_c^2}{\tau_Q}\sigma\right)^{1/4}.
\label{equ:kccond}
\end{equation}
Now, we will have to find out what this means in terms of the number
density of vortices. By definition, the modes with $k\le k_c$ freeze
out, and those with $k\ge k_c$ equilibrate. This magnetic field is
squeezed into vortices, but let us ignore that for a moment, and
assume that all modes with $k\ge k_c$ simply disappear. This means
that the magnetic field configuration is a superposition of waves with
wavelength longer than $1/k_c$, and if we look at the system at
distances shorter than that, we simply see a uniform magnetic field. 
Eventually, all this flux turns into vortices (see
Fig.~\ref{fig:trapping}c).  
If there
was initially a flux $\Phi(R)$ through a circle of radius
$R\lsim 1/k_c$, it will form $\Phi(R)/\Phi_0$
vortices, all with the same sign.
The number density $n$ can then be calculated by dividing this by the
area $\pi R^2$ of the circle.

On the other hand, if $R\gsim 1/k_c$, the Fourier modes with $k\ge
k_c$ only affect the spatial distribution of the flux inside
the curve, but do not contribute to the net flux. Consequently, the
final flux $\Phi(R)$ after the transition is the same as it was in the
symmetric phase before the transition. This can be calculated from
$G(k)$ 
or estimated using physical arguments, 
and the result is~\acite{Hindmarsh:2000kd}
\begin{equation}
\Phi(R)\approx \sqrt{TR}.
\label{equ:fluxR}
\end{equation}

If we now choose $R=1/k_c$, we are on the borderline of these limiting
cases. We can say that the flux is given by Eq.~(\ref{equ:fluxR}),
and that the
whole flux is turned into vortices of equal sign. Therefore, we know
what $\Phi(1/k_c)$ is, and we can calculate the number density of
vortex lines,
\begin{equation}
n\approx \left.\frac{\Phi(R)/\Phi_0}{\pi R^2}\right|_{R=1/k_c}
\approx \sqrt{e^2Tk_c^3}.
\label{equ:trappred}
\end{equation}
Using Eq.~(\ref{equ:kccond}), we can write this as
\begin{equation}
n\approx e^{7/4}T_c^{5/4}\left(\frac{\sigma}{\tau_Q}\right)^{3/8}.
\end{equation}

\begin{figure}[t]
\center
\epsfig{file=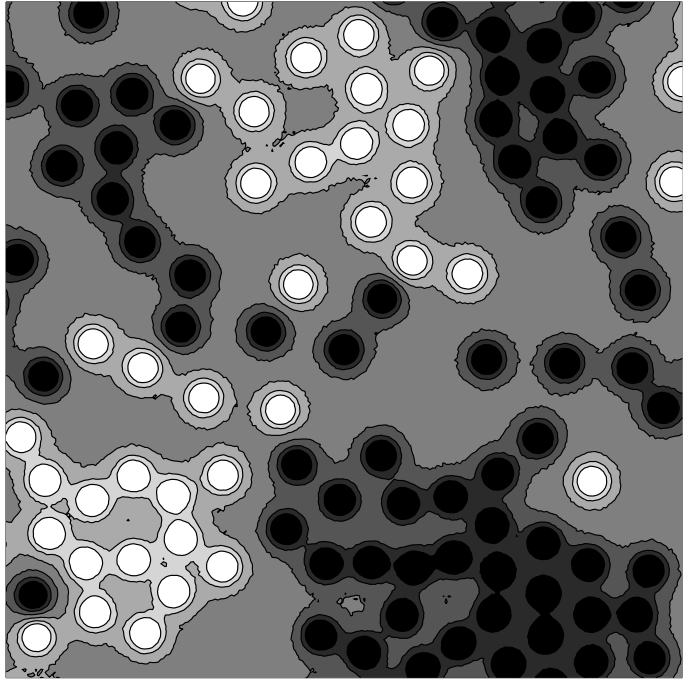,width=9cm}
\caption{
\label{fig:stephens}
A snapshot from a numerical simulation of the Abelian Higgs model. 
It shows the magnetic field configuration after a quench into the
broken phase.
Black and
white correspond to magnetic fields pointing up and down,
respectively. The clustering of vortices is obvious. 
(From Stephens et al.~\cite{Stephens:2001fv}).
}
\end{figure}

\begin{figure}[t]
\center
\epsfig{file=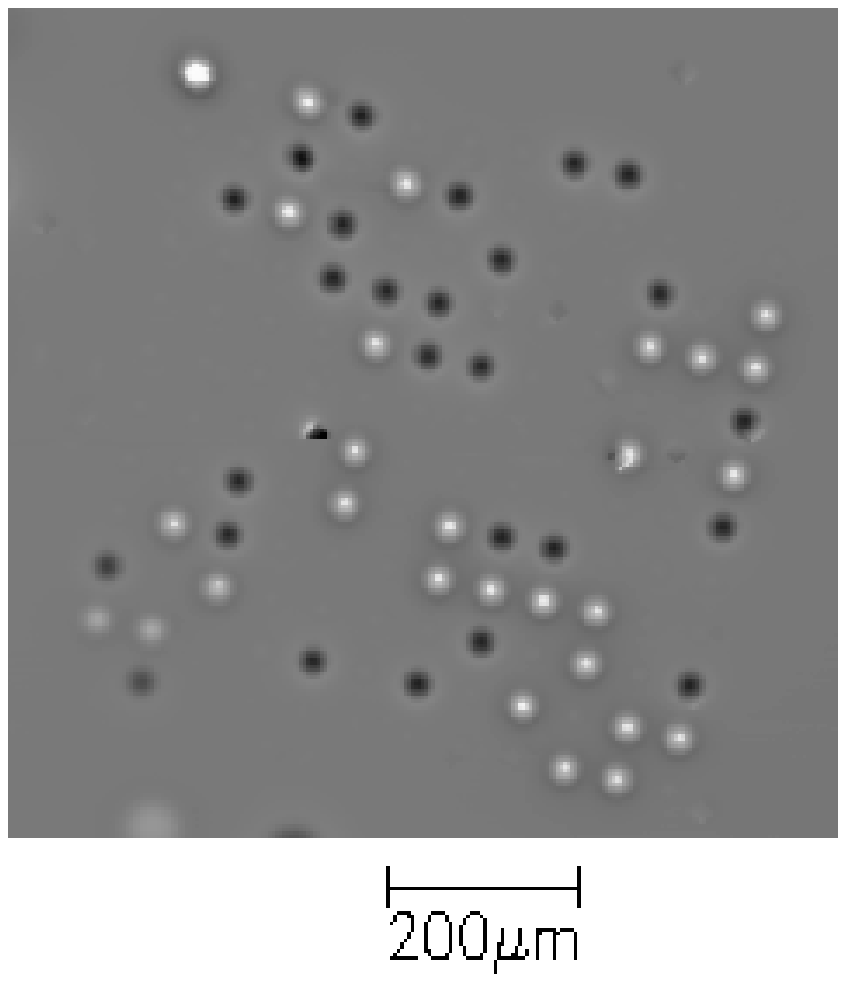,width=9cm}
\caption{
\label{fig:kirtley}
The distribution of magnetic flux in an array of superconducting
rings after a rapid phase transition.
The similarity with Fig.~\ref{fig:stephens} is striking,
and the authors report that their results show that thermal fluctuations
play an important role. The result is still not a conclusive proof of 
the theory discussed in Section~\ref{sect:trapping}, because a ring
array is in many ways different from a superconducting film or indeed a
three-dimensional space.
(From Kirtley et al.~\cite{Kirtley})
}
\end{figure}

The two-dimensional analogue of this result has been tested in
numerical simulations~\acite{Stephens:2001fv}, and 
seems to work well (see Fig.~\ref{fig:stephens}).
There have also been attempts to test the result experimentally with
superconductors (see Fig.~\ref{fig:kirtley})~\acite{Carmi,Kirtley,Maniv}.
However, the actual experiments
are usually done with
two-dimensional films.
This is not equivalent to the fully two-dimensional setup used in the
simulations,
because in the
experimental setup, the magnetic field extends outside the
film. Therefore, the magnetic field behaves as in three dimensions and
the scalar field as in two dimensions. This situation was discussed
briefly
in
Ref.~\acite{Hindmarsh:2000kd}, and a more detailed analysis is given in
Ref.~\acite{kibble}.

However, just like the Kibble mechanism, this scenario can also give
predictions that are independent of dynamical parameters such as the
conductivity. As we saw, the flux $\Phi(R)$ through a circle with
$R=1/k_c$ turns into vortices of equal sign, and therefore, if it is
greater than $\Phi_0$, we should see clusters of equal-sign vortices.
Based on Eq.~(\ref{equ:trappred}), we would expect roughly
$N_{\rm cl}\approx (e^2T/k_c)^{1/2}$ 
vortices per cluster. Because we can make $k_c$
arbitrarily small by cooling the system slower, we should be able to
make these clusters if we can control the cooling rate. 

If $N_{\rm cl}$ is large enough, it is much more likely that a
randomly picked vortex is inside a cluster than near it
boundary. Therefore, if we calculate the quantity $\sigma_W(R)$
defined in Section~\ref{sect:kibble}, all the vortices contributing to
$N_W(R)$ are of the same sign as long as $R$ is less than the cluster
size $1/k_c$. Therefore $N_W(R)$ is proportional to its area, and
consequently, 
\begin{equation}
\sigma_W(R)\propto R^4,\quad\text{(Flux trapping)}
\end{equation}
at short distances. In other words, $\nu_\sigma=1$.

Furthermore, if we can measure both the number density $n$ and the
cluster size $N_{\rm cl}$, the combination $nN_{\rm cl}^3$ should be
independent of $k_c$ and can therefore be used to test the scenario.
On the other hand, if one wants to learn more about the dynamics of the
phase transition, one would measure that combination $n/N_{\rm cl}$,
which would give the freeze-out scale $k_c$.

The
result in Eq.~(\ref{equ:trappred})
is proportional to the square root of temperature and
would therefore vanish as $T$ goes to zero. This is important, because
the transition that ended hybrid inflation would have taken place at
zero temperature. Would defect formation then have happened according
to the Kibble mechanism, or would the gauge fields still have played
some role? In the above discussion, we have only considered classical
thermal fluctuations, and it might be possible that quantum
fluctuations play a similar role at zero temperature. More work is
needed to find out whether that is really the case.

I have argued recently in Ref.~\acite{Rajantie:2002dw} that the
formation of magnetic monopoles in the GUT phase transition may take
place in a somewhat similar way. There are several extra
complications, though. In particular, as was already mentioned, there
may actually be no true phase transition at all, just a smooth
crossover from one phase to another. One may then ask whether
monopoles are formed at all.

It is, however, easy to see that the theory cannot stay in equilibrium
if it is cooled at a constant rate. There are no long-range
correlations in the symmetric phase of the theory, and it is generally
known that any correlation length has to be of the order of
$(g^2T)^{-1}$ or shorter~\acite{Linde:ts}, 
where $g$ is again the gauge coupling
constant. 
On the other hand, at a very low temperature
we should recover normal electrodynamics, which has a massless
photon. Correspondingly, the magnetic field has an infinite
correlation length. 

Admittedly, it is not obvious
what ``magnetic field'' means in the symmetric phase, but no matter
how we define it, we have a correlation length $\xi_{\bf B}$ that starts from a
microscopic value $(g^2T)^{-1}$ and grows to an infinite
value. Analogously with Kibble's argument, we can say that $\xi_{\bf B}$
cannot grow arbitrarily fast, because causality alone restricts its
growth rate to be less than the speed of light. It is therefore
inevitable that if we keep on cooling the system, it falls out of
equilibrium at some point and the magnetic correlation length freezes
to a finite value, which we denote by $\hat\xi_{\bf B}$.

The next key observation is that 
the finiteness of the correlation length is due to
screening. This is analogous to the Debye screening of
electric field by electric charges~\acite{plasmabook}, 
and therefore it implies a
presence of magnetic charges. Thus, because the correlation length is
necessarily finite due to causality, magnetic monopoles must have been
formed. 

These arguments can be made more quantitative 
by noting that one can define a magnetic charge that is conserved even
in the symmetric phase. This charge behaves in a similar way to the
magnetic field in the Abelian Higgs model. In the symmetric phase,
there are long-wavelength thermal fluctuations, which should disappear
in the broken phase, but again, causality implies that fluctuations
whose wavelength exceeds some critical value do not have time to
decay. They freeze out and form vortices. This critical wavelength
turns out to be exactly the value $\hat\xi_{\bf B}$ 
to which the correlation length of the
magnetic field freezes.
One finds that the
number density of monopoles formed at the transition would be 
roughly~\acite{Rajantie:2002dw}
\begin{equation}
n_M\approx \sqrt{\frac{g^2T}{\hat\xi_{\bf B}^5}}.
\end{equation}
As for vortices, one also find that if $\hat\xi_{\bf B}$
is long enough, clusters of $\sqrt{g^2T\hat\xi_{\bf B}}$ monopoles are
formed.

So far, we have been discussing continuous (or smooth) phase
transitions, but discontinuous first-order transitions are possible 
both in Abelian and non-Abelian theories, and in general they take
place if the Higgs self coupling constant $\lambda$ is less than the
square of the gauge coupling constant.

In a first-order transition, bubbles of the broken phase are
nucleated, and they grow and coalesce. We can safely assume that there
are no defects inside the bubbles. Because magnetic field and
charge are conserved in Abelian and non-Abelian cases, respectively,
they will be trapped in the gaps between the bubbles. Therefore, we
would expect that monopoles are formed in well-localized clusters.
In the Abelian case,
vortices with multiple windings are stable if there is a first-order
transition, and therefore we would
expect thick and massive vortex lines, which carry much higher fluxes
than in the continuous case~\acite{Kibble:1995aa}.

\section{Defects in cosmology}
\label{sect:cosmology}

There is no conclusive
evidence for the existence of topological defects in the
universe, but there is strong evidence against certain types of
topological defects. First of all, domain walls are ruled out unless
they, for some reason, have an extremely low energy
density~\acite{VilenkinShellard}. The
reason for this is that the mass of a domain wall is proportional to
its area and is therefore enormous for a wall that extends through the
whole universe.

Global strings or monopoles could arise from a spontaneous breakdown
of a continuous global symmetry, but as we saw in
Section~\ref{sect:global}, that
would predict the existence of massless Goldstone bosons. No such
particles have been observed, but they may still exist if they only interact
extremely weakly with other forms of matter.

Otherwise, we are left to consider gauged defects. Cosmic strings
have been studied intensively, because it was thought they could
explain the primordial density perturbations that gave rise to the
large-scale structure of the universe and the temperature fluctuations
of the cosmic microwave background (CMB) radiation. However, recent
CMB measurements~\acite{Spergel:2003cb} 
show that this is not the case, and that these
perturbations probably originated in quantum fluctuations during
inflation. This observation does not by any means rule out the
existence of cosmic strings, and people are still looking for
evidence for or against them in the CMB radiation and in other
astronomical observations. One way of finding strings is by
gravitational lensing. Because strings are so massive, they would bend the 
light coming from galaxies behind them and therefore we would see 
two copies of the same galaxy. A possible such
observation of a cosmic string has been reported recently~\cite{Sazhin:2003cp}.

In principle, magnetic monopoles would be a natural candidate for the
missing dark matter, which is known to make up about 85\% of all the
matter in the universe~\acite{Spergel:2003cb}. They have been
searched for in experiments, but with no
success~\acite{Ambrosio:2002qq}. This seems to rule out the possibility
of superheavy magnetic monopoles with GUT scale masses being the main
component of the dark matter. There is also a more severe problem:
The number of magnetic monopoles produced at the
GUT phase transition has been estimated to be so high that their mass
would have caused the universe to collapse very soon after the
transition~\acite{Zeldovich:wj,Preskill:1979zi}. This is known as the {\it monopole
problem}. 

In the standard Big Bang scenario with no inflation, we would assume
that the temperature was initially around the Planck scale
$10^{19}$~GeV. When the universe was expanding and cooling down, it
went though the GUT transition at around $10^{16}$~GeV.
Because all GUTs predict the existence
of magnetic monopoles and they would have been formed at this
transition, this scenario suffers from the
monopole problem.

The monopole problem is solved by inflation if the reheat
temperature was low enough so that the GUT symmetry was never
restored. So far, there are very few other observational constraints on the
reheat temperature, and more or less the only other constraint comes
from the matter-antimatter asymmetry. 
There is overwhelming observational evidence that there is practically
no antimatter but a significant density of matter in the
universe. It seems impossible to generate this asymmetry after
the electroweak phase transition~\acite{Shaposhnikov:vs}, 
which took place at around
$100$~GeV, and therefore the reheat temperature should be high enough
for the electroweak symmetry to be restored.
This gives the rough bounds,
\begin{equation}
100~\text{GeV}~\lsim ~T_{\rm rh}~\lsim ~10^{16}~\text{GeV}.
\label{equ:reheat_bounds}
\end{equation}
If there were other phase transitions, 
cosmic strings may have been produced provided that
the reheat temperature was above the corresponding critical temperature.

There could, however, be extra complications. 
First of all, it is not at all clear, how well the scenarios that
assume an equilibrium initial state can really describe phase
transitions that take place in a rapidly expanding universe only a
short time after the end of inflation. The assumption of thermal
equilibrium seems particularly questionable at long distances, between
points that 
have only barely come within each other's sight.

In many
inflationary models, reheating can be a very violent non-equilibrium
process, in which case it is often called 
{\it preheating}~\acite{Kofman:1994rk}. 
It may then be possible to produce topological
defects~\acite{Kofman:1995fi,Khlebnikov:1998sz,Kasuya:1998td,Tkachev:1998dc} 
even if the reheat temperature is well below the critical
temperature. In some cases, it is possible to describe this as a
``non-thermal'' symmetry 
restoration. It is not known how well the
scenarios discussed in Sections~\ref{sect:kibble} and
\ref{sect:trapping} 
can describe defect
formation from preheating. In any case, it seems that we may have to
rethink the upper bound in Eq.~(\ref{equ:reheat_bounds}). The same
applies to the lower bound, because similar
effects may also create the matter-antimatter 
asymmetry~\acite{Krauss:1999ng,Garcia-Bellido:1999sv,Rajantie:2000nj}.

There is also a third possibility in
hybrid inflation (\ref{equ:hybrid}). 
Because inflation ends at a breakdown of a
symmetry, there is always a phase transition at the end of inflation,
and that would typically lead to defect 
formation~\acite{Linde:1993cn,Felder:2001kt,Copeland:2002ku}. 
If we want to avoid
domain walls and Goldstone bosons, the waterfall field $\chi$ will have
to be a multi-component field with a gauge symmetry. Because
the universe is at zero temperature at the time of the phase
transition, it is not yet clear if the scenario discussed in
Section~\ref{sect:trapping} is applicable.

\section{Conclusions}
Theoretically, it seems plausible that topological defects should have
been formed in the early universe. If signs of them are observed,
that will give an enormous boost to
our understanding of the universe,
but the failure to find them can also tell us many things. Most 
significantly, the absence of the high number of monopoles predicted by
the theory led to the development of the theory of inflation, which
has become the standard paradigm in cosmology because of other supporting
observations. The apparent absence of defects can also help us
constrain certain parameters of the inflationary models, which would
otherwise be undetermined. This will
become more important in the future, when we start to home in on a
particular class of inflationary models using data from other
observations. It also has significance in particle physics, because
it constrains possible high-energy extensions of the Standard Model of
particle physics.

For all this it is important to have a good understanding of how defects
are formed. Currently, there are theories that describe this process
in transitions that start from thermal equilibrium in either global or
gauge field systems. In the global case, some of the predictions have
been confirmed in condensed matter experiments, and it seems that the
gauge theory predictions will be tested with superconductors in the
future. 

However, the transitions in the early universe may have taken place
far away from thermal equilibrium, and much more work is still needed
before we can really claim that we understand how possible
cosmological defects would have
been formed.

\section*{Acknowledgements}
The author was supported by PPARC, and also partly by Churchill
College and the ESF COSLAB program.


\begin{thebibliography}{0}

\abibitem{Peierls:fg}
R.~Peierls,
Contemp.\ Phys.\  {\bf 33} (1992) 221.

\abibitem{Kibble:1976sj}
T.~W.~B.~Kibble,
J.\ Phys.\ A {\bf 9} (1976) 1387.

\abibitem{Zurek:qw}
W.~H.~Zurek,
Nature {\bf 317} (1985) 505.

\abibitem{VilenkinShellard}
A.~Vilenkin and E.P.S.~Shellard,
{\it Cosmic Strings and Other Topological Defects}
(Cambridge University Press, Cambridge, 1994).

\abibitem{Vachaspati:1998vc}
T.~Vachaspati,
Contemp.\ Phys.\  {\bf 39} (1998) 225.

\abibitem{Gangui:2003uu}
A.~Gangui,
http://arXiv.org/abs/astro-ph/0303504

\abibitem{Gill:1998}
A.J.~Gill,
Contemp.\ Phys.\  {\bf 39} (1998) 13.

\abibitem{Gleiser:1998kk}
M.~Gleiser,
Contemp.\ Phys.\  {\bf 39} (1998) 239.

\abibitem{Rajantie:2001ps}
A.~Rajantie,
Int.\ J.\ Mod.\ Phys.\ A {\bf 17} (2002) 1.

\abibitem{gaugebook}
D.~Bailin and A.~Love,
{\it Introduction to Gauge Field Theory},
(Institute of Physics Publishing, Bristol, 1995).

\abibitem{superconductors}
M.~Tinkham, {\it Introduction to Superconductivity}, (McGraw-Hill, New
York, 1996).

\abibitem{Georgi:sy}
H.~Georgi and S.~L.~Glashow,
Phys.\ Rev.\ Lett.\  {\bf 32} (1974) 438.

\abibitem{Kephart:2001ix}
T.~W.~Kephart and Q.~Shafi,
Phys.\ Lett.\ B {\bf 520} (2001) 313.

\abibitem{Davis:1986nr}
R.~L.~Davis,
Phys.\ Rev.\ D {\bf 35} (1987) 3705.

\abibitem{Nielsen:cs}
H.~B.~Nielsen and P.~Olesen,
Nucl.\ Phys.\ B {\bf 61} (1973) 45.

\abibitem{'tHooft:1974qc}
G.~'t Hooft,
Nucl.\ Phys.\ B {\bf 79} (1974) 276.

\abibitem{Polyakov:ek}
A.~M.~Polyakov,
JETP Lett.\  {\bf 20} (1974) 194
[Pisma Zh.\ Eksp.\ Teor.\ Fiz.\  {\bf 20} (1974) 430].

\abibitem{Spergel:2003cb}
D.~N.~Spergel {\it et al.},
http://arXiv.org/abs/astro-ph/0302209.

\abibitem{Guth:1980zm}
A.~H.~Guth,
Phys.\ Rev.\ D {\bf 23} (1981) 347.

\abibitem{lindebook}
A.~D.~Linde, {\it Particle Physics and Inflationary Cosmology},
(Harwood Academic, Chur, 1990).

\abibitem{Copeland:gs}
E.~J.~Copeland,
Contemp.\ Phys.\  {\bf 34} (1993) 303.

\abibitem{Linde:1993cn}
A.~D.~Linde,
Phys.\ Rev.\ D {\bf 49} (1994) 748.

\abibitem{Jackiw:cv}
R.~Jackiw,
Phys.\ Rev.\ D {\bf 9} (1974) 1686.

\abibitem{Kirzhnits:ut}
D.~A.~Kirzhnits and A.~D.~Linde,
Phys.\ Lett.\ B {\bf 42} (1972) 471.

\abibitem{Dolan:qd}
L.~Dolan and R.~Jackiw,
Phys.\ Rev.\ D {\bf 9} (1974) 3320.

\abibitem{Hasenbusch:1998gh}
M.~Hasenbusch, K.~Pinn and S.~Vinti,
Phys.\ Rev.\ B {\bf 59} (1999) 11471.

\abibitem{Hasenbusch:1999cc}
M.~Hasenbusch and T.~Torok,
J.\ Phys.\ A {\bf 32} (1999) 6361.

\abibitem{Laguna:1997cf}
P.~Laguna and W.~H.~Zurek,
Phys.\ Rev.\ D {\bf 58} (1998) 085021.

\abibitem{Elitzur:im}
S.~Elitzur,
Phys.\ Rev.\ D {\bf 12} (1975) 3978.

\abibitem{Kirzhnits:1976ts}
D.~A.~Kirzhnits and A.~D.~Linde,
Annals Phys.\  {\bf 101} (1976) 195.

\abibitem{Hart:1996ac}
A.~Hart, O.~Philipsen, J.~D.~Stack and M.~Teper,
Phys.\ Lett.\ B {\bf 396} (1997) 217.

\abibitem{Hendry}
P.~C.~Hendry {\it et al.},
Nature {\bf 368} (1994) 315.

\abibitem{Dodd}
M.~E.~Dodd {\it et al.},
Phys.~Rev.~Lett.~{\bf 81} (1998) 3703.

\abibitem{Ruutu:1996qz}
V.~M.~Ruutu {\it et al.},
Nature {\bf 382} (1996) 334.

\abibitem{Bauerle}
C.~B\"auerle {\it el al.},
Nature {\bf 382} (1996) 332.

\abibitem{TurokNature}
I.~Chuang, R.~Durrer, N.~Turok and B.~Yurke,
Science {\bf 251} (1991) 1336.

\abibitem{Bowick:1994rz}
M.~J.~Bowick, L.~Chandar, E.~A.~Schiff and A.~M.~Srivastava,
Science {\bf 263} (1994) 943.

\abibitem{Digal:1998ak}
S.~Digal, R.~Ray and A.~M.~Srivastava,
Phys.\ Rev.\ Lett.\ {\bf 83} (1999) 5030.

\bibitem{Ducci1999}
S.~Ducci, P.~L.~Ramazza, W.~Gonzalez-Vinas and F.~T.~Arecchi,
Phys.\ Rev.\ Lett.\ {\bf 83} (1999) 5210.

\bibitem{Casado2001}
S.~Casado, W.~González-Viñas, H.~Mancini and S.~Boccaletti,
Phys.\ Rev.\ E {\bf 63} (2001) 057301.

\abibitem{Hindmarsh:2000kd}
M.~Hindmarsh and A.~Rajantie,
Phys.\ Rev.\ Lett.\  {\bf 85} (2000) 4660.

\abibitem{Stephens:2001fv}
G.~J.~Stephens, L.~M.~Bettencourt and W.~H.~Zurek,
Phys.\ Rev.\ Lett.\  {\bf 88} (2002) 137004.

\abibitem{Carmi}
R.~Carmi and E.~Polturak,
Phys.\ Rev.\ B {\bf 60} (1999) 7595.

\abibitem{Kirtley}
J.~R.~Kirtley, C.~C.~Tsuei and F.~Tafuri,
Phys.\ Rev.\ Lett.\ {\bf 90} (2003) 257001.

\abibitem{Maniv}
A.~Maniv, E.~Polturak and G.~Koren,
http://arXiv.org/abs/cond-mat/0304359.

\abibitem{kibble}
T.~W.~B.~Kibble and A.~Rajantie,
http://arXiv.org/abs/cond-mat/0306633.

\abibitem{Rajantie:2002dw}
A.~Rajantie,
Phys.\ Rev.\ D {\bf 68} (2003) 021301.

\abibitem{Linde:ts}
A.~D.~Linde,
Phys.\ Lett.\ B {\bf 96} (1980) 289.

\abibitem{plasmabook}
R.~J.~Goldsten and P.~H.~Rutherford, 
{\it Introduction to Plasma Physics},
(Institute of Physics Publishing, Bristol, 1995).

\abibitem{Kibble:1995aa}
T.~W.~B.~Kibble and A.~Vilenkin,
Phys.\ Rev.\ D {\bf 52} (1995) 679.

\abibitem{Sazhin:2003cp}
M.~Sazhin {\it et al.},
Mon.\ Not.\ Roy.\ Astron.\ Soc.\ {\bf 343} (2003) 353.

\abibitem{Ambrosio:2002qq}
M.~Ambrosio {\it et al.}  [MACRO Collaboration],
Eur.\ Phys.\ J.\ C {\bf 25} (2002) 511.

\abibitem{Zeldovich:wj}
Y.~B.~Zeldovich and M.~Y.~Khlopov,
Phys.\ Lett.\ B {\bf 79} (1978) 239.

\abibitem{Preskill:1979zi}
J.~P.~Preskill,
Phys.\ Rev.\ Lett.\  {\bf 43} (1979) 1365.

\abibitem{Shaposhnikov:vs}
M.~E.~Shaposhnikov,
Contemp.\ Phys.\  {\bf 39} (1998) 177.

\abibitem{Kofman:1994rk}
L.~Kofman, A.~D.~Linde and A.~A.~Starobinsky,
Phys.\ Rev.\ Lett.\  {\bf 73} (1994) 3195.

\abibitem{Kofman:1995fi}
L.~Kofman, A.~D.~Linde and A.~A.~Starobinsky,
Phys.\ Rev.\ Lett.\  {\bf 76} (1996) 1011.

\abibitem{Khlebnikov:1998sz}
S.~Khlebnikov, L.~Kofman, A.~D.~Linde and I.~Tkachev,
Phys.\ Rev.\ Lett.\  {\bf 81} (1998) 2012.

\abibitem{Kasuya:1998td}
S.~Kasuya and M.~Kawasaki,
Phys.\ Rev.\ D {\bf 58} (1998) 083516.

\abibitem{Tkachev:1998dc}
I.~Tkachev, S.~Khlebnikov, L.~Kofman and A.~D.~Linde,
Phys.\ Lett.\ B {\bf 440} (1998) 262.

\abibitem{Krauss:1999ng}
L.~M.~Krauss and M.~Trodden,
Phys.\ Rev.\ Lett.\  {\bf 83} (1999) 1502.

\abibitem{Garcia-Bellido:1999sv}
J.~Garcia-Bellido, D.~Y.~Grigoriev, A.~Kusenko and M.~E.~Shaposhnikov,
Phys.\ Rev.\ D {\bf 60} (1999) 123504.

\abibitem{Rajantie:2000nj}
A.~Rajantie, P.~M.~Saffin and E.~J.~Copeland,
Phys.\ Rev.\ D {\bf 63} (2001) 123512.

\abibitem{Felder:2001kt}
G.~N.~Felder, L.~Kofman and A.~D.~Linde,
Phys.\ Rev.\ D {\bf 64} (2001) 123517.

\abibitem{Copeland:2002ku}
E.~J.~Copeland, S.~Pascoli and A.~Rajantie,
Phys.\ Rev.\ D {\bf 65} (2002) 103517.

\end{thebibliography}
\end{document}